\documentclass[final,3p,times]{elsarticle}
\usepackage{amssymb}
\usepackage{algorithm}
\usepackage{algpseudocode}
\usepackage{amsmath}
\usepackage{xurl}
\usepackage{tabularx} % Include the tabularx package for more flexible tables
\usepackage{booktabs}

\usepackage{xcolor,colortbl}
\definecolor{grayshade}{RGB}{230,230,230}

\usepackage[english]{babel}
\usepackage{amsthm}

\usepackage{gensymb}

\usepackage{hyperref}
\usepackage[all]{hypcap} % hyperref jump to figure instead of captions
\hypersetup{colorlinks=true,allcolors=blue}

\journal{Computer Methods in Applied Mechanics and Engineering}

\begin{document}

\begin{frontmatter}

\title{Neurodevelopmental disorders modeling using isogeometric analysis, dynamic domain expansion and local refinement}

% Use letters for affiliations, numbers to show equal authorship (if applicable) and to indicate the corresponding author
\author[a]{Kuanren Qian}
\author[b]{Genesis Omana Suarez}
\author[b]{Toshihiko Nambara}
\author[b]{Takahisa Kanekiyo}
\author[a]{Ashlee S. Liao}
\author[a,c,d]{Victoria A. Webster-Wood}
\author[a,c]{Yongjie Jessica Zhang\corref{cor1}}

\address[a]{Department of Mechanical Engineering, Carnegie Mellon University, 5000 Forbes Ave, Pittsburgh, PA 15213, USA}
\address[b]{Department of Neuroscience, Mayo Clinic, 4500 San Pablo Road, Jacksonville, FL 32224, USA}
\address[c]{Department of Biomedical Engineering, Carnegie Mellon University, 5000 Forbes Ave, Pittsburgh, PA 15213, USA}
\address[d]{McGowan Institute for Regenerative Medicine, University of Pittsburgh, 450 Technology Drive, Pittsburgh, PA 15219, USA}

\cortext[cor1]{Corresponding author. \\ \hspace*{2em} \textit{E-mail address:} \href{mailto:jessicaz@andrew.cmu.edu}{jessicaz@andrew.cmu.edu} (Y. J. Zhang).}

\begin{abstract}
Neurodevelopmental disorders (NDDs) have arisen as one of the most prevailing chronic diseases within the US. 
Often associated with severe adverse impacts on the formation of vital central and peripheral nervous systems during the neurodevelopmental process, NDDs are comprised of a broad spectrum of disorders, such as autism spectrum disorder, attention deficit hyperactivity disorder, and epilepsy, characterized by progressive and pervasive detriments to cognitive, speech, memory, motor, and other neurological functions in patients.
However, the heterogeneous nature of NDDs poses a significant roadblock to identifying the exact pathogenesis, impeding accurate diagnosis and the development of targeted treatment planning. 
A computational NDDs model holds immense potential in enhancing our understanding of the multifaceted factors involved and could assist in identifying the root causes to expedite treatment development.
To tackle this challenge, we introduce optimal neurotrophin concentration to the driving force and degradation of neurotrophin to the synaptogenesis process of a 2D phase field neuron growth model using isogeometric analysis to simulate neurite retraction and atrophy.
The optimal neurotrophin concentration effectively captures the inverse relationship between neurotrophin levels and neuron survival, while its degradation regulates concentration levels. 
Leveraging dynamic domain expansion, the model efficiently expands the domain based on outgrowth patterns to minimize degrees of freedom.
Based on truncated T-splines, our model simulates the evolving process of complex neurite structures by applying local refinement adaptively to the cell/neurite boundary. 
Furthermore, a thorough parameter investigation is conducted with detailed comparisons against neuron cell cultures in experiments, enhancing our fundamental understanding of the possible mechanisms underlying NDDs.

\end{abstract}

\begin{keyword}
Neuron growth \sep Neurodevelopmental disorders \sep Phase field method \sep Isogeometric analysis \sep Dynamic domain expansion \sep Truncated T-splines \sep Local refinement
 %% PACS codes here, in the form: \PACS code \sep code
%% MSC codes here, in the form: \MSC code \sep code
%% or \MSC[2008] code \sep code (2000 is the default)
\end{keyword}

\end{frontmatter}

\section{Introduction}

Neurodevelopmental disorders (NDDs), often associated with impairments during the neuron developmental process, pose persistent complications that have a profound and long-lasting impact on patients such as attention deficit hyperactivity disorder (ADHD) and autism spectrum disorder~\cite{thapar2017neurodevelopmental, tager1999neurodevelopmental}. 
A wide range of potential factors behind the etiology of NDDs are gaining recognition recently, with emerging research highlighting their significant impact on the pathophysiology of these conditions~\cite{fujitani2021pathophysiological, yamamoto2021genomic}. 
Investigations into the functional roles of these factors have revealed some potential influences on NDDs progression, indicating that they play crucial roles during the neurodevelopmental process~\cite{dugger_pathology_2017, brown_neurodegenerative_2005}. 
Moreover, emerging evidence suggests that certain factors involved during the neurodevelopmental process may play protective roles, potentially preventing the onset of NDDs~\cite{connor_role_1998, berg_new_1984}.
However, the lack of comprehensive disorder studies of the specific biological functions or biophysical processes highlights a significant gap in the current field.
Furthermore, NDDs often manifest through complex neuron morphological transformations, such as beading, retraction, and atrophy. 
All of these behaviors add complexity to a thorough neurological study and pose challenges to therapeutic targeting and planning~\cite{elliott_motor_1996, datar2019roles}. 
Analyzing the complex morphological changes observed in NDDs and understanding the biological mechanisms that drive these conditions are crucial for developing effective therapeutic strategies. 
Given the level of complexity associated with the sophisticated processes involved, unconventional approaches are necessary to enhance our understanding of the NDDs process. 
As such, there is a pressing need for a comprehensive computational model to study NDDs and unravel the intricate neurodevelopmental process.
These models will be indispensable for shedding light on the fundamental causes and complexities of NDDs, consequently enabling the development of precise and potent treatments~\cite{budday2015physical, wang2023multi}. 

In computational neuroscience, significant progress has been made in mathematically modeling neurodevelopmental processes, including initial neurite outgrowth~\cite{hentschel_instabilities_1994}, axon differentiation~\cite{krottje_mathematical_2007, pearson_mathematical_2011} and axon guidance~\cite{aeschlimann_biophysical_2001}. 
Phenomenological models have provided insights into a wide range of neurodevelopmental processes such as filopodia~\cite{goodhill_predicting_2004}, external repulsive cues~\cite{maskery_growth_2004}, stochastic mechanisms~\cite{koene_netmorph_2009}, generalized neurite characteristics for modeling morphology~\cite{cuntz_one_2010, donohue_comparative_2008}, and incorporating interactions with surrounding substrates~\cite{torben-nielsen_context-aware_2014}. 
However, these models focus on phenomenological outcomes, often overlooking underlying biophysics~\cite{eberhard_neugen_2006, van_ooyen_independently_2014}.
Integrating comprehensive biophysical mechanisms into neuron growth models comes with significant computational costs and numerical instabilities~\cite{otoole_physical_2008, graham_mathematical_2006}, which are particularly aggravated in scenarios like intracellular material transport~\cite{li_deep_2021} and traffic jams in complex 3D structures~\cite{li2023isogeometric}. 
Despite these complications, ongoing efforts are underway to enhance and validate computational models with biophysical phenomena from experimental observations. 
Among different approaches, some noteworthy work includes the utilization of phase field techniques for neuron growth modeling~\cite{takaki_phase_field_2015}, exploration of neurotrophin interactions~\cite{nella2022bridging}, and the development of biophysically coupled phase field neuron growth models~\cite{qian_modeling_2022}. 
There are also models that integrate neurite morphometric features and bridge the gap between theoretical and experimental neuroscience~\cite{qian2023biomimetic, liao_quantitative_2022}. 
Considering the complexity and large variety of the neurodevelopmental process~\cite{van_ooyen_modeling_2003}, robust computational approaches that tackle high-order equations on complex geometries are crucial for advancing our understanding of neurodevelopmental processes and paving the way toward effective targeted therapeutic interventions.
This raises the need for isogeometric analysis (IGA) with high-fidelity spline modeling techniques that provide the necessary flexibility and precision~\cite{zhang_challenges_2013, zhang_geometric_2018}, particularly through domain expansion and localized refinements to accurately capture the dynamic and detailed evolution of neuron growth during development and degeneration in response to damage or pathology.

Non-uniform rational B-splines are widely adopted~\cite{piegl1996nurbs, gordon_b-spline_1974} and were initially chosen as the foundational basis for IGA~\cite{hughes_isogeometric_2005, cottrell2009isogeometric}.
However, B-splines lack local refinement support, an important feature for our NDDs study. 
To address this limitation, T-splines have been developed to work with IGA and support the local refinements required for efficient and accurate analysis~\cite{casquero_isogeometric_2016}.
Local refinement in T-splines is achieved with T-junctions that are analogous to hanging nodes in conventional finite element methods (FEM), which break down global tensor product structures~\cite{sederberg2003t, sederberg2004t}.
This adaptability particularly benefits IGA because it lowers the degrees of freedom (DOFs) necessary while preserving the exact representations. 
In addition to local refinement, T-splines maintain essential properties of B-splines, including non-negativity and partition of unity. 
Thus, they extend the functionality of B-splines through local refinements while preserving the underlying mathematical properties. 
These properties make T-splines highly desirable in a wide range of problems, therefore leading to the development of T-splines into various forms such as analysis-suitable T-splines~\cite{scott2012local}, LR-splines~\cite{dokken2013polynomial, johannessen2014isogeometric}, modified T-splines~\cite{kang2013modified, wei2022analysis}, and weighted T-splines~\cite{liu2015weighted, liu2015handling}. 
Hierarchical approaches such as
% as HB-splines~\cite{vuong2011hierarchical, bornemann2013subdivision}, PHT-splines~\cite{deng2008polynomial}, THB-splines~\cite{giannelli2012thb, pawar2018dthb3d_reg}, 
PHT-splines~\cite{deng2008polynomial}, 
hierarchical analysis-suitable T-splines~\cite{evans2015hierarchical}, and truncated hierarchical Catmull–Clark subdivision~\cite{wei2015truncated, wei2016extended} have also been used for local refinement, with additional advancements in adaptive refinement techniques~\cite{pawar2016adaptive}.
%These developments highlight that not all T-splines are suitable for analysis. 
%Only a specific subset, analysis-suitable T-splines, are adapted to meet the stringent requirements of computational analyses~\cite{scott2012local, buffa2010linear, li2012linear}. 
The capability of T-splines is further showcased in modeling heterogeneous solids~\cite{li2020trivariate}, additive manufacturing analysis~\cite{li2019slicing}, and arbitrary-degree T-splines for IGA of Kirchhoff-Love shells~\cite{casquero2017arbitrary}. 
%Additional applications include hybrid-degree weighted T-splines~\cite{liu2016hybrid}, analysis-suitable unstructured T-splines~\cite{casquero2016isogeometric}, and integration into unstructured mesh environments~\cite{casquero2020seamless}. 
In particular, the applications of T-splines in intracellular material transport modeling~\cite{li2023isogeometric, li2022modeling, li2022modeling_1} further underline the versatility and potential of IGA and truncated T-splines in computational neuroscience.

%As shown in EPA data, the prevalence of NDDs among the youth necessitates sophisticated computational models to investigate the mechanisms behind neuron developmental processes and NDDs. 
To this end, we propose a novel IGA phase-field NDDs model coupling neuron growth with complex biophysics processes, 
built upon dynamic domain expansion, local refinement, and the Portable, Extensible Toolkit for Scientific Computation (PETSc)~\cite{petsc-user-ref,petscsf2022} for Message Passing Interface (MPI) parallelization.
%our model can achieve scalable parallelizations utilizing high-performance computation resources.
% to handle large DOFs.
%Leveraging locally refined truncated T-splines, the model can conduct efficient and accurate neuron growth simulations. 
%In addition, we implemented dynamic domain expansion to minimize computational costs. 
With this model, we can investigate how neurons behave
% shrink or deteriorate 
under NDDs and the underlying biophysics processes. 
The main contributions include:
\begin{itemize}
    \item Development of a PETSc-based IGA phase field NDDs model. The NDDs model simulates complex neuron growth and disorder behaviors on truncated T-splines, leveraging PETSc for efficient parallel processing;
    \item Introduction of optimal neurotrophin concentration into the driving force and degradation of neurotrophin into the synaptogenesis process of the phase field model to simulate neurite retraction and atrophy. 
    The optimal neurotrophin concentration effectively captures the inverse relationship between neurotrophin levels and neuron survival, and its degradation regulates concentration levels, providing hypotheses into NDD mechanisms;
    \item Dynamic domain expansion that optimizes computational focus by expanding the domain based on neurite growth and interface-based local refinements that refine at evolving interface to preserve accuracy while minimizing computational load;
%     The model leverages dynamic domain expansion to optimize computational focus by expanding the domain based on neurite growth. It also xxxx local refinements at evolving interface, preserving accuracy while minimizing computational load;
    \item Comprehensive NDDs model parameter study utilizing the computational model to investigate neurite retraction and atrophy, providing new insights into possible factors and biophysical mechanisms underlying NDDs; and
    % and enhancing our understanding of their biophysical mechanisms.
    \item  Detailed qualitative validation through comparisons of the external-cue guided computational NDDs model with experimental healthy human induced pluripotent stem cell (iPSC)-derived neurons and rat hippocampus neurons undergoing cell death and neurite fragmentation.
\end{itemize}

The rest of this paper is organized as follows. 
Section~\ref{sec: overview} outlines our NDDs model structure.
Section~\ref{sec: IGA NG model} introduces our IGA-based phase field model for NDDs. 
Section~\ref{sec: truncated t-splines} reviews the truncated T-splines and introduces our phase field variable interface-based local refinements. 
Section~\ref{sec: domain expansion} walks through our dynamic domain expansion algorithm and interpolation process. 
Section~\ref{sec: experimental culture} documents the experimental procedures for culturing human iPSCs-derived neurons and rat hippocampal neurons.
Section~\ref{sec: results} explains the parallelized computational model and showcases neuron growth and disorders simulation results with experimental comparisons.
Section~\ref{sec: conclusion} concludes the NDDs model and discusses the potential future directions.

\section{Algorithm overview}
\label{sec: overview}

\begin{figure}[h]
    \centering
    \includegraphics[width=\textwidth]{
        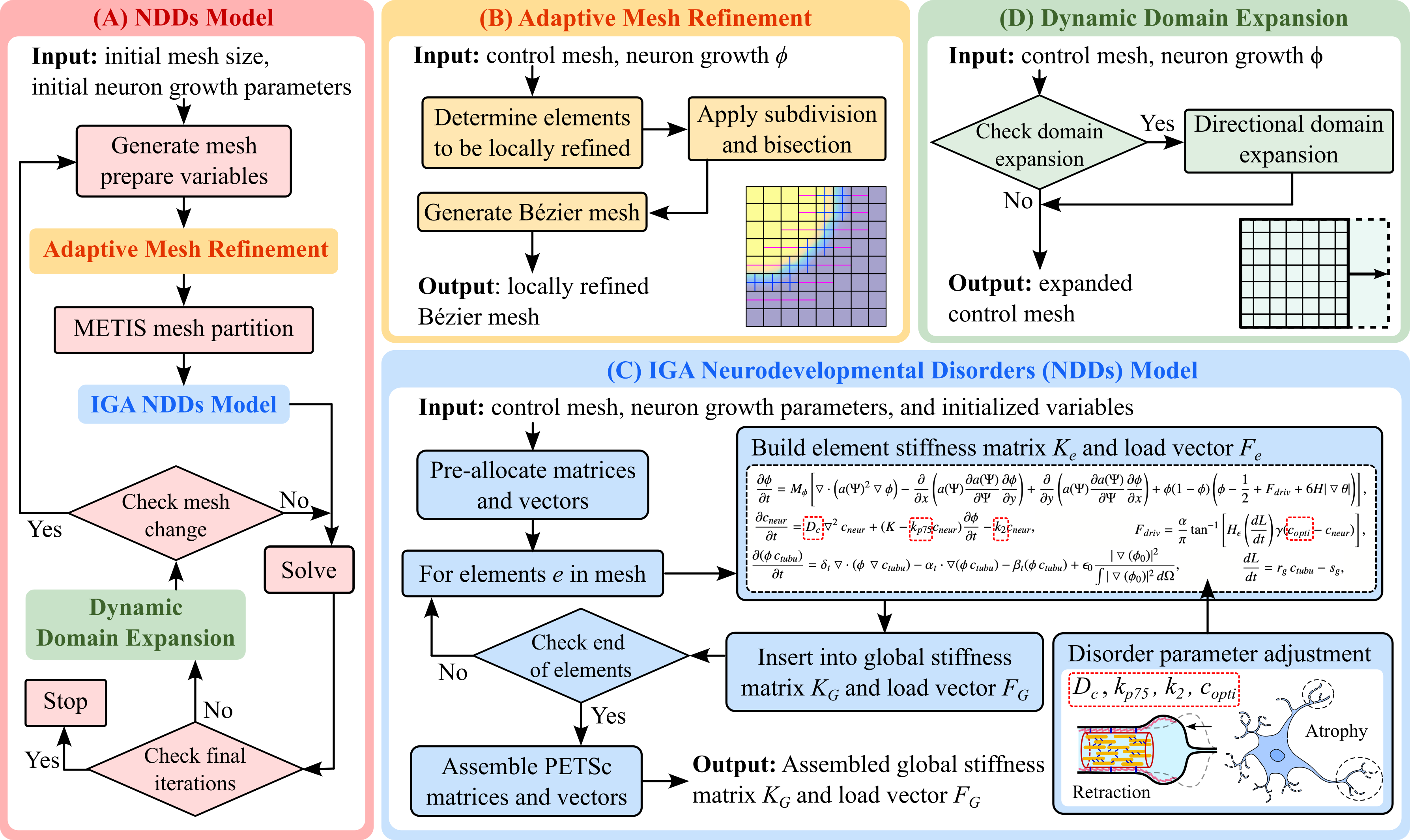
    }
    \caption{
        Flow chart of the NDDs modeling pipeline.
        (A) The overall pipeline that conducts NDDs modeling.
        (B) Adaptive mesh refinement module that locally refines the mesh based on neuron outgrowth.
        (C) The IGA NDDs model simulates disorders using the phase field method, PETSc, and truncated T-splines. Parameters in red dashed boxes are selected to study their effects on NDDs.
        (D) Dynamic domain expansion module that directionally expands domain based on neurites near the domain boundary.
%        (For better interpretation of annotations in the figure, readers are referred to the web version of this article.)
    }
    \label{fig: flowchart}
\end{figure} 

The flowchart illustrates our NDDs modeling workflow (Fig.~\ref{fig: flowchart}). 
The model starts with predefined domain size and neuron growth parameters, which are used to generate the initial control mesh and variables (Fig.~\ref{fig: flowchart}A).
Then, the adaptive mesh refinement module (Fig.~\ref{fig: flowchart}B) is called to generate truncated T-splines.
This module checks for control mesh elements that require local refinements based on phase field variable $\phi$ value and selectively refines elements located on the $\phi$ interface, achieving sufficient resolution while keeping DOFs minimum.
The relevant elements ID are selected and passed if refinement is needed to generate locally refined B\'ezier mesh~\cite{wei2017truncated}. 
Subsequently, mesh partitioning software METIS is called to partition the generated mesh for optimizing the computational load distribution across multiple processors~\cite{karypis1998fast}. 
%With the partitioned mesh assigned to each processor, matrices needed for the simulation are preallocated and prepared based on the current iteration number for efficient parallelization. 
The model then builds and solves all nonlinear and linear systems using  PETSc (Fig.~\ref{fig: flowchart}C)~\cite{petsc-user-ref}.
Afterward, the model checks whether the final iterations have been reached. 
If not, the model checks for neurites near domain boundaries and expands the domain as needed using the dynamic domain expansion module (Fig.~\ref{fig: flowchart}D). 
If the mesh changes, the model regenerates the truncated T-splines to ensure that all B\'ezier elements have up-to-date control point information.
The above steps are iterated until the final iteration is reached.

\section{IGA-based phase field NDDs modeling}
\label{sec: IGA NG model}

IGA and phase field methods are robust and powerful numerical methods for modeling complex engineering problems~\cite{gomez_accurate_2014, schillinger_isogeometric_2015}. 
IGA is a high-order numerical method that can capture the exact smooth representation of the geometry by eliminating the discretization needed by conventional FEM~\cite{hughes_isogeometric_2005}. 
The phase field method, on the other hand, specializes in tackling evolving boundaries such as crack propagation and dendritic solidification~\cite{takaki2014phase}. 
Considering that NDDs are neuron growth processes with retracting cell boundaries, they are essentially an interface evolution problem.
The convergence of IGA and phase field is highly effective in accurately simulating neurite morphological transformations.
Utilizing these two techniques, an IGA-based phase field framework was introduced to depict the complex stages of healthy neuron growth by incorporating intracellular concentration~\cite{qian_modeling_2022, qian2023biomimetic}. 
This model accounts for intracellular transport during the growth, enables the differentiation of the longest neurite into an axon, and models various growth dynamics across multiple stages to simulate the behavior of growth cones at neurite tips based on neurite morphometric features~\cite{liao2023semi}. 
The proposed NDDs model extends upon it, consisting of five equations:
\begin{enumerate}
    \item Phase field governing equation that captures the neuron morphological transformation with phase field variable;
    \item Intracellular tubulin transport equation that models the effect of tubulin on neurite elongations;
    \item Competitive tubulin consumption equation that captures the consumption of tubulin at neurite tips;
    \item Synaptogenesis equation that captures the effect of neurotrophin particles supporting interface evolution. 
    We include the effect of neurotrophin degradation that regulates its concentration level to model NDDs; and
    \item Driving force equation that couples the effect of tubulin and neurotrophin back to the phase field equation. 
    We introduce optimal neurotrophin concentration to model the inverse relationship between neurotrophin level and neuron survival. 
    This inverse relationship adjusts the driving force magnitude and, therefore, drives the phase field interface evolution for NDDs modeling.
\end{enumerate}

\textbf{Phase field governing equation.} To model NDDs in the context of phase field, we treat the neuron domain $\Omega$ as a binary phase field $\phi$, where phase ``1'' indicates the neuron and phase ``0'' is the extracellular environment. 
The phase field interface evolution is achieved by constantly solving for energy minimization at interfaces.
Starting with the simplified free energy functional for the phase field model as
\begin{equation}
    E_{free} = \int_\Omega \left( E_{chem} + E_{grad} + E_{doub} \right) d\Omega,
    \label{eqn: free energy functional equation}
\end{equation} where $E_{chem}$ is the chemical free energy density
%, often expressed using a monotonically increasing function $q(\phi) = \phi^2(1-\phi)^2$ to interpolate bulk energy between solid and liquid.
that expresses the bulk energy relationship,
$E_{grad}$ is the gradient energy density~\cite{takaki_phase_field_2015}, and
$E_{doub}$ is the double-well function that describes the barrier in free energy density as a function of the phase field parameter $\phi$ with two stable states representing the two phases of the system.
Then, following the modified Allen–Cahn equation~\cite{takaki2014phase, allen1979microscopic}, we have
\begin{equation}
    \frac{\partial \phi}{\partial t} = -M_{\phi} \frac{\delta E_{free}}{\delta \phi}, \label{eqn: time derivation equation}\\
\end{equation} 
where the functional derivative $\frac{\delta E_{free}}{\delta \phi}$ is derived based on Eqn.~\ref{eqn: free energy functional equation}:
\begin{equation}
    \frac{\delta E_{free}}{\delta \phi} = \frac{\partial E_{chem}}{\partial \phi} - \nabla \cdot \frac{\partial E_{grad}}{\partial (\nabla \phi)} + \frac{\partial E_{doub}}{\partial \phi}, \label{eqn: dFdPhi equation}
\end{equation}
with
\begin{align}
E_{chem} &= p(\phi) f_s + (1 - p(\phi)) f_l, \label{eqn: f_chemical}\\
E_{grad} &= \frac{a(\Psi)^2}{2} |\nabla \phi|^2, \label{eqn: f_gradient} \\
E_{doub} &= Wq(\phi). \label{eqn: f_double}
\end{align}
Following approach taken in existing literature~\cite{takaki2014phase,takaki2007phase}, $E_{chem}$ interpolates a monotomically increasing bulk free energy relationship between solid $f_S$ and liquid $f_L$ with $p(\phi) = \phi^3(10 - 15\phi + 6\phi^2)$.
$E_{grad}$ is defined using the anisotropy gradient coefficient $a(\Psi)$ to account for thin neurite morphology, $E_{doub}$ is defined using double well function $q(\phi) = \phi^2(1-\phi)^2$.
$W$ is the energy barrier magnitude.
% By substituting Eqns.~\ref{eqn: f_chemical}-\ref{eqn: f_gradient} into Eqn~\ref{eqn: dFdPhi equation}, we obtain
% \begin{equation}
% \frac{\delta F}{\delta \phi} = 30\phi^2 (1-\phi)^2 (f_s - f_l) + 2W\phi (1-\phi) (1-2\phi) - a^2 \nabla^2 \phi.
% \end{equation} 
% Then, we can derive Eqn.~\ref{eqn: time derivation equation} as
% \begin{equation}
%     \frac{\partial \phi}{\partial t} = -M_{\phi} , \label{eqn: xxxx}\\
% \end{equation} 
Substituting Eqns.~\ref{eqn: dFdPhi equation}-\ref{eqn: f_double} into Eqn~\ref{eqn: time derivation equation}, we obtain
\begin{equation}
    \frac{\partial \phi}{\partial t} = M_{\phi} \left[ 
    \bigtriangledown \cdot \left(a(\Psi)^2\nabla \phi \right)
    % a(\psi)^2 \nabla^2 \phi 
    + 4W\phi(1-\phi) \left(\phi - \frac{1}{2} + \frac{15}{2W} \phi(1-\phi) (f_L - f_S)\right)\right]. \label{eqn: initial phase}
\end{equation}
The anisotropic gradient coefficient $a(\Psi)$~\cite{takaki_phase_field_2015, eggleston2001phase} is defined as
\begin{equation}
a(\Psi) = 
\begin{cases} 
a_1(\Psi) = \frac{\bar a}{1 + \xi} \left[1 + \xi \cos \left( k(\Psi - \theta) \right) \right] & \text{for } \pi i + \theta_m \leq \Psi - \theta \leq \pi(i+1) - \theta_m, \\
a_2(\Psi) = \frac{a_1(\theta_m) \cos(\Psi - \theta)}{\cos \theta_m} & \text{for } \pi i - \theta_m < \Psi - \theta < \pi i + \theta_m. \label{eqn: anisotropic gradient coefficient equation}
% \alpha_1(\psi) = \frac{\bar{\alpha}}{1 + \gamma} \left[ 1 + \gamma \cos \left( k \left( \psi - \frac{2\pi\theta}{k} \right) \right) \right], \quad \alpha_2(\psi) = \frac{\alpha_1(\theta_m + \theta) \cos(\psi - \theta)}{\cos(\theta_m)}
\end{cases}
\end{equation}
where $\bar a$ is a scaling constant, $\xi$ the anisotrophy strength, $k$ is the anisotrophy mode, $i$ is 0 and 1, and $\theta_m$ is the missing orientation under high anisotrophy~\cite{eggleston2001phase,takaki2006two}.
$\theta$ is the orientation that denotes the variation in neurite elongation direction~\cite{takaki_phase_field_2015}. 
We also include the driving force term $F_{driv}$ to couple with the orientation field~\cite{qian_modeling_2022, ren_controllable_2018}.
Then, the phase field governing equation becomes
% \begin{equation}
%     \frac{\partial \phi}{\partial t} 
%     = M_{\phi} \left[\bigtriangledown\cdot \left(a(\Psi)^2\bigtriangledown \phi \right) 
%     - \frac{\partial} {\partial{x}}\left(a(\Psi) \frac{\partial a(\Psi)}{\partial \Psi} \frac{\partial \phi}{\partial y}\right) 
%     + \frac{\partial} {\partial{y}}\left(a(\Psi) \frac{\partial a(\Psi)}{\partial \Psi} \frac{\partial \phi}{\partial x}\right) 
%     + \phi(1-\phi) \left(\phi - \frac{1}{2} + F_{driv} + 6H |\bigtriangledown \theta| \right) \right], \label{eqn: phase field equation}
% \end{equation}
\begin{equation}
    \frac{\partial \phi}{\partial t} 
    = M_{\phi} \left[\nabla \cdot \left(a(\Psi)^2 \nabla \phi \right) 
    - \frac{\partial} {\partial x}\left(a(\Psi) \frac{\partial a(\Psi)}{\partial \Psi} \frac{\partial \phi}{\partial y}\right) 
    + \frac{\partial} {\partial y}\left(a(\Psi) \frac{\partial a(\Psi)}{\partial \Psi} \frac{\partial \phi}{\partial x}\right) \right. \left. + \phi(1-\phi) \left(\phi - \frac{1}{2} + F_{driv} + 6H |\nabla \theta| \right) \right], \label{eqn: phase field equation}
\end{equation} where $M_{\phi}$ denotes the mobility coefficient of the phase field equation, 
%$a(\Psi)$ is the anisotropy coefficient, 
$F_{driv}$ is the driving force for the evolution of the phase field variable. 
$H$ is a constant value, and the term $6H |\nabla \theta|$ is introduced to disrupt symmetry~\cite{ren_controllable_2018}. 
% in dendritic growth~\cite{ren_controllable_2018}. 

\textbf{Intracellular tubulin transport equation.}
Tubulin is the building block of microtubules in cells, transported to neurite tips via active transport and diffusion and necessary to support neurite elongations~\cite{mclean2004continuum,mclean2004mathematical}.
Developing based on 1D tubulin model~\cite{graham2006dynamics}, we can model the effect of intracellular tubulin transport as
\begin{equation}
    \frac{\partial (\phi \,c_{tubu})}{\partial t} =  \delta_t \bigtriangledown\cdot \: (\phi \, \nabla c_{tubu}) - \mathbf{\alpha}_{t} \cdot \bigtriangledown (\phi \, c_{tubu}) - \beta_{t} (\phi \, c_{tubu}) + \epsilon_0 \frac{|\nabla(\phi_0)|^2}{\int{|\nabla(\phi_0)|^2} \, d \Omega}, 
    \label{eqn: tubulin equation}
\end{equation}
where $\delta_t$ is the rate at which tubulin diffuses, $\alpha_t$ is the active transport coefficient for tubulin, $\beta_t$ is the decay coefficient for tubulin, and $\epsilon_0 \frac{|\nabla(\phi_0)|^2}{\int{|\nabla(\phi_0)|^2} d\Omega}$ represents constant tubulin production. 
$\phi_0$ is the initial phase field, and $\epsilon_0$ is coefficient for tubulin production. 
%This equation consists of tubulin diffusion, transport, decay, and production in the neuron growth process. 

\textbf{Competitive tubulin consumption equation. }
Tubulin concentration is then used to calculate competitive tubulin consumption at neurite tips, which is a crucial factor that determines neurite outgrowth:
\begin{equation}
\frac{dL}{dt} = r_{g} \, c_{tubu} - s_{g}, 
\label{eqn: dLdt equation}
\end{equation}
where $r_g$ and $s_g$ are the tubulin assembly and disassembly rate~\cite{mclean2004mathematical}, and $\frac{dL}{dt}$ reflects the effect of dynamic tubulin consumption balance that is crucial for neurite extension~\cite{mclean2004continuum, van_ooyen_competition_2001}.

\textbf{Synaptogenesis equation.}
Drawing inspiration from previous studies on heat conduction in dendritic solidification~\cite{kobayashi_modeling_1993}, we adopt the concept to simulate the diffusion of neurotrophin concentration in neurons~\cite{nella2022bridging}. 
In order to model NDD behaviors, we incorporate the effect of neurotrophin into the model.
Since neurotrophin diffusion is crucial for guiding growth cone and neurite pathways during synaptogenesis progress~\cite{song1997camp}, we utilize the evolution of the $\phi$ interface, denoted by $\frac{\partial \phi}{\partial t}$, as the neurotrophin source generation in the synaptogenesis equation.
The survival or death of neurons is influenced by the differential binding of neurotrophins to receptors such as $p75NTR$ receptors, a type of transmembrane proteins located near the neuron growth cone~\cite{bamji1998p75, barrett2000p75}.
Its balance is discovered to be a critical factor in neuron survival following injury~\cite{meeker2014dynamic, meeker2015p75}. 
Since $p75NTR$ receptors are proven to be fast-diffusive monomers and bind to neurotrophin concentration~\cite{marchetti2019fast}, we introduce a degradation $k_{p75}c_{neur}$ into the source generation in the synaptogenesis equation to model the associated degradation~\cite{nella2022bridging}. 
We obtain
\begin{equation}
    \frac{\partial c_{neur}}{\partial t} = D_c \nabla^2 c_{neur} + (K - k_{p75}c_{neur})\frac{\partial \phi}{\partial t} - k_2 c_{neur}, \label{eqn: synaptogenesis equation}
\end{equation}
where $c_{neur}$ is the neurotrophin concentration, $D_c$ is the diffusion coefficient, and $K$ is the latent neurotrophin. 
$k_{p75}$ is the degradation rate when binding to $p75NTR$ receptors, and $k_2 c_{neur}$ is the sink term for the degradation~\cite{nella2022bridging, krewson1996transport}.
Because $c_{neur}$ is vital in neurite outgrowths and directly supports the $\phi$ interface balance and evolution, the degradation to $c_{neur}$ introduced by $k_{p75}$ and $k_2$ can lead to abnormal neurite morphological transformations associated with NDDs.

\textbf{Driving force equation.}
In neurons, neurotrophin concentrations are a determining factor in neuron survival. 
An optimal amount of neurotrophin is necessary for the survival of neurons~\cite{lu2005yin}. 
Lower levels typically support neuron survival, while higher concentrations can negate this effect. 
This inverse relationship between neurotrophin concentrations and cell survival is particularly notable at higher concentrations, where an observed survival-promoting effect at lower concentrations is reversed~\cite{piontek1999neurotrophins}.
To model the aforementioned inverse relationship between neurotrophin concentrations and cell survival considering the intricate balance of neurotrophin concentration~\cite{huang2001neurotrophins}, we introduce $c_{opti}$ to
% The effects of the tubulin equation (Eqn.~\ref{eqn: tubulin equation}) and synaptogenesis equation (Eqn.~\ref{eqn: synaptogenesis equation}) are then coupled back to the model through the driving force equation defined as
the driving force equation and obtain
\begin{equation}
    F_{driv} = \frac{\alpha}{\pi} \tan^{-1}\left[H_\epsilon\left(\frac{dL}{dt}\right) \gamma (c_{opti}-c_{neur})\right],
\label{eqn: driving force equation}
\end{equation} 
where 
%$E$ is the energy term driving interface changes. 
%$\frac{\alpha}{\pi}$ is a scaling coefficient, $H_\epsilon$ is a Heaviside step function, $\gamma$ is the interfacial energy constant, and $c_{opti}$ is the optimal concentration for neuron survival. 
% $F_{driv}$ is the driving force term that drives cell growth at the phase interface. 
$\frac{\alpha}{\pi}$ is a scaling coefficient, $H_\epsilon$ is a Heaviside step function, and $\gamma$ is the interfacial energy constant.
% \textcolor{blue}{Explain how $c_{opti}$ is related to NDDs.}
By introducing $c_{opti}$, Eqn.~\ref{eqn: driving force equation} can effectively model the inverse relationship between $c_{neur}$ and neuron survival.
% $c_{opti} - c_{neur}$ models the inverse effect of neurotrophin concentration on the driving force. 
% As $c_{neur}$ changes relative to $c_{opti}$, the difference directly controls the strength of the driving force. 
When the magnitude of $c_{neur}$ exceeds $c_{opti}$, the effect of $c_{neur}$ is reversed, leading to the $\phi$ interface retractions.
% \textcolor{blue}{align writing with contribution point}

\vspace{2mm}
In the implementation, we use IGA to solve the phase field governing equation (Eqn. \ref{eqn: phase field equation}) concurrently with the intracellular tubulin transport equation (Eqn.~\ref{eqn: tubulin equation}) and the synaptogenesis equation (Eqn.~\ref{eqn: synaptogenesis equation}), coupled through the driving force equation (Eqn.~\ref{eqn: driving force equation}) with competitive tubulin consumption equation (Eqn.~\ref{eqn: dLdt equation}).

\section{Truncated T-splines and local refinement}
\label{sec: truncated t-splines}

Since the $\phi$ domain is mostly stable with values consistently at ``0'' and ``1'' except at the evolving interface, we utilize truncated T-splines to apply local refinements at $\phi$ interface within the IGA framework to accelerate the computation.
This approach prioritizes computational efforts on the areas undergoing evolutions over iterations, maintaining high accuracy while avoiding extensive mesh refinement by substantially decreasing the number of DOFs. 
This leads to quicker solver iterations and increased computational efficiency.

\subsection{Review of truncated T-spline}
T-spline is developed to lift the limitations of the uniform control grid B-splines~\cite{sederberg2003t}.
%a generalized extension of the B-spline, developed to address the restrictions posed by the uniform control grid B-splines require~\cite{sederberg2003t}.
B-splines rely on a uniform, rectangular grid, which has limited capability when modeling intricate geometries and often requires more control points. 
A univariate B-spline of order $p$ is defined on a non-decreasing sequence of real numbers, $u_i$, which are used to build a knot vector $\bar U = \{u_1, u_2, \cdots u_{n+p+1}\}$, where $n$ is the number of basis functions and $p$ is the order of the B-spline~\cite{piegl1996nurbs}. 
The basis function $N_{i,p}(u)$ is defined recursively as follows:
\begin{equation}
    N_{i,0}(u) = 
    \begin{cases}
        1 & \text{if } u_i \leq u < u_{i+1}, \\
        0 & \text{otherwise},
    \end{cases}
    \label{eqn: basis function piecewise linear}
\end{equation}
\begin{equation}
    N_{i,p}(u) = \frac{u-u_i}{u_{i+p}-u_i} N_{i,p-1}(u) + \frac{u_{i+p+1}-u}{u_{i+p+1}-u_{i+1}} N_{i+1,p-1}(u),
    \label{eqn: basis function recusive}
\end{equation}
where $N_{i,0}(u)$ is a piece-wise linear basis function, $N_{i,p}(u)$ is the B-spline basis function of degree $p$, recursively calculated based on combinations of two degree $(p-1)$ basis functions over the knot vector $\bar U$. 
Given a set of control points $P = {\{P_{i}\}}_{i=1}^n$, the B-spline curve $C(u)$ is defined as:
\begin{equation}
    C(u) = \sum_{i=0}^n N_{i,p}(u) P_i, \quad 0 \leq u \leq 1.
\end{equation}
%where $p$ is order of the spline, and $P$ is the set of control points.

We can obtain a T-spline control mesh by introducing T-junctions into a quadrilateral mesh in the physical domain.
Each edge in the mesh is assigned a parametric value, known as the \textit{knot interval}, that requires an equal cumulative sum across the opposite edge of the element to ensure continuity and uniform parameterization across the mesh.
T-spline basis functions differ from B-splines in that they are defined using local knot vectors, determined by shooting rays across the T-mesh in the parametric directions~\cite{wei2017truncated}. 
A ray is extended to intersect with the mesh to determine the local knot vector for a particular point. 
The intersection coordinates are collected and arranged in ascending order to form the local knot vector. 
This allows the T-spline basis functions to be directly influenced by the local topology of the T-mesh, enhancing their flexibility and ability to accurately represent geometry with less restriction than the global knot vectors used in B-splines.
%Utilizing the Cox-de Boor formula~\cite{cottrell2009isogeometric}, uni-variate B-spline basis functions are derived for each local knot vector, denoted as $N_{U}(u)$ and $N_{V_A}(v)$. 
T-splines enhance these capabilities by supporting local refinement via knot insertion~\cite{lyche1985knot}. 
%In T-spline refinement, the knot insertion technique enhances the spline space locally~\cite{lyche1985knot}. 
Given a univariate T-spline basis function $N_{U} (u)$ based on an initial knot vector $U = \{u_{1}, u_{2}, u_{3}, u_{4}, u_{5}\}$, we can extend $U $ by adding $n$ knots, where
%$n$ belongs to all positive integers 
$n \in \mathbb{Z}^+$, to obtain an enlarged and refined knot vector $U' = \{u'_{1} = u_{1}, u'_{2}, \ldots, u'_{n+5} = u_{5}\} \supset U$. Here $U'_{i} = \{u'_{i}, u'_{i+1}, \ldots, u'_{i+4}\}~(i = 1, \ldots, n+1)$ are enriched local knot vectors. 
The locally refined basis function $N_{U'} (u)$ is then defined as:
\begin{equation}
    N_{U'} (u) = \sum_{i=1}^{n+1} c_i N_{U'_{i}} (u),
\end{equation}
where $c_i$ is the refinement coefficients from knot insertion~\cite{lyche1985knot}, and $N_{U'_{i}} $ are the active children of $N_{U'} (u)$.
A bi-variate T-spline basis function, $B(u, v) = N_{U}(u) N_{V}(v)$, is the tensor product of two uni-variate functions, and its refinement is achieved by refining $N_{U}(u)$ and $N_{V}(v)$~\cite{wei2017truncated}.

While locally refined T-splines offer significant flexibility for geometric modeling, they can pose challenges for analysis due to difficulties in maintaining linear independence due to overlapping basis functions across different refinement levels. 
Analysis-suitable T-splines were introduced to address these problems with standard T-splines, such as ensuring linear independence and maintaining the partition of unity~\cite{scott2012local, buffa2010linear} by limiting intersections except under specific conditions that ensure analysis suitability.
Truncation is an approach that has been applied to T-splines to enhance their robustness and flexibility by allowing certain intersections, except for face-face types~\cite{wei2017truncated}. 
This approach, initially implemented in truncated hierarchical B-splines~\cite{giannelli2012thb, giannelli2014strongly}, has been extended to truncated hierarchical Catmull-Clark subdivision~\cite{wei2015truncated, wei2016extended}.
To preserve the partition of unity and geometric integrity, the truncation mechanism mitigates overlapping basis functions by selectively excluding redundant contributions from active children functions in the refinement hierarchy. 
For a T-mesh $T$ and its refined counterpart $T'$ from truncated T-spline quadtree subdivision, a bi-variate partially refined basis function $B_i(u, v)$ needs to exclude redundant children $B_j'(u, v)$ to avoid overlapping influences of basis functions~\cite{wei2017truncated}. 
The truncated basis function, denoted as $ trunB_i(u, v)$, is defined as
\begin{equation}
    trunB_i(u, v) = B_i(u, v) - \sum_{j \neq i, B_j' \in B_i} c_{ij} B_j'(u, v),
\end{equation}
where $c_{ij}$ are coefficients determined in the knot insertion process, $B_i(u, v)$ is the parent basis functions, and $B_j'(u, v)$ represents the set of partially refined children basis functions being discarded.
This ensures that $trunB_i(u, v)$ forms a partition of unity while preserving local refinement effects.
The truncated T-spline surface, $S(u, v)$, is defined as
\begin{equation}
    S(u, v) = \frac{\sum_{i=0}^n trunB_{ij}(u, v) P_{ij} }{\sum_{i=0}^n trunB_{ij}(u, v) },
\end{equation}
where $P_{ij}$ are the control points and $trunB_{ij}(u, v)$ are the truncated T-spline basis functions. 
Note that $trunB_{ij}(u, v)$ at any particular point sums up to 1 to ensure the property of partition of unity~\cite{liunurbs}, which makes truncated T-splines particularly effective for accurate computational analysis with complex geometries. 
We refer readers to~\cite{wei2017truncated} for an in-depth discussion on truncated T-splines.

\subsection{$\phi$ interface-based local refinements}

\begin{figure}[h]
    \centering
    \includegraphics[width=\textwidth]{
        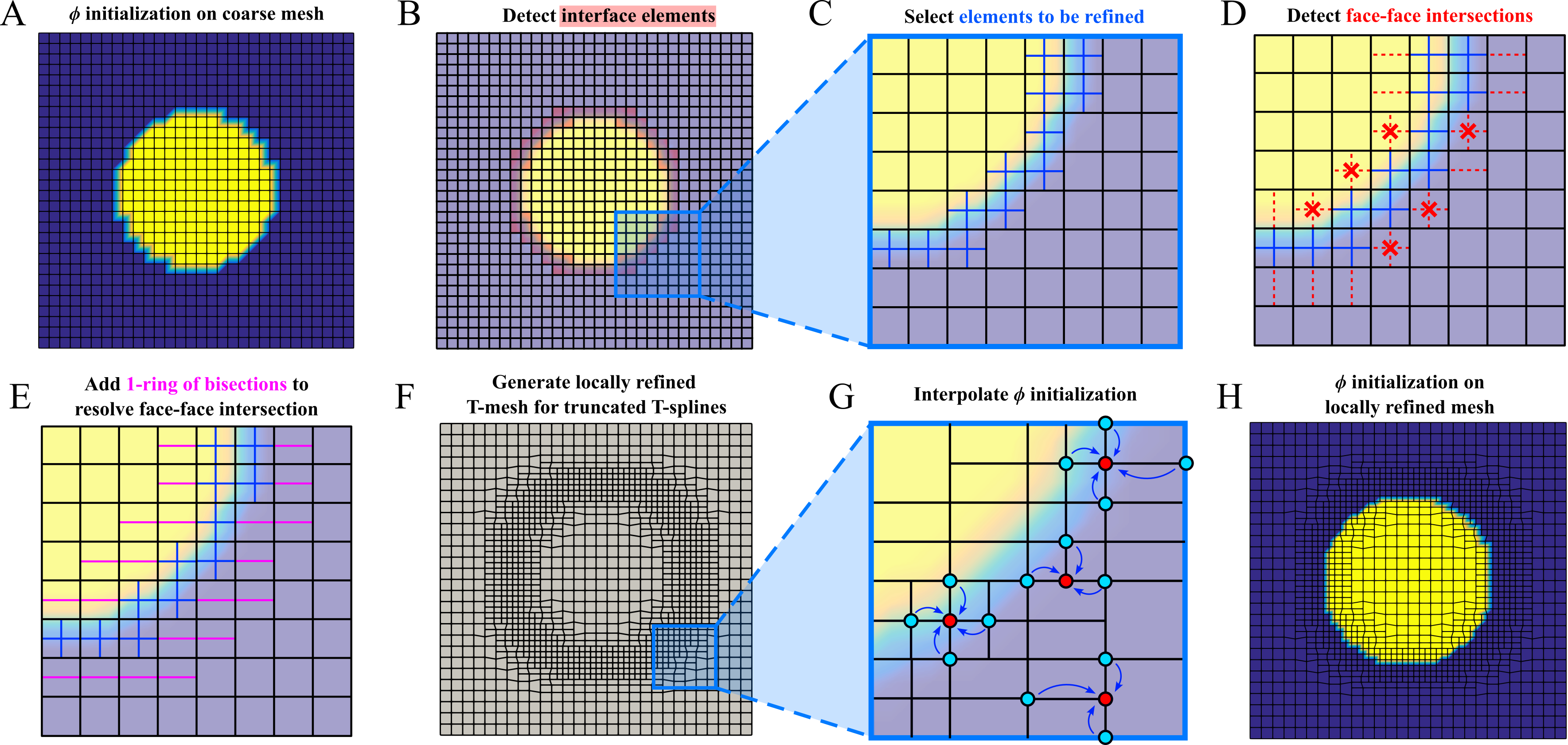
    }
    \caption{
        Local refinements using truncated T-splines. 
        (A) Phase field variable $\phi$ initialization on a coarse mesh. 
        (B) Identifying elements at the $\phi$ interface.
        (C) Interface elements are locally refined with subdivision. 
        (D) Detecting face-face intersections created by face extension from subdivisions. 
        (E) Applying 1-ring of bisections to resolve these face-face intersections and maintaining the integrity of the mesh for truncated T-splines.
        (F) Generating locally refined T-mesh for truncated T-splines with enhanced accuracy in regions of interest. 
        (G) Interpolating $\phi$ initialization from the coarse control mesh to the locally-refined mesh.
        (H) The locally refined $\phi$ initialization.
        %(For better interpretation of annotations in the figure, readers are referred to the web version of this article.)
    }
    \label{fig: truncated t-spline}
\end{figure} 

%Beginning with the initial phase field variable $\phi$, which is shown on a uniform control mesh (Fig.~\ref{fig: truncated t-spline}A), we can identify the $\phi$ interface and extract associated element ID. 
To solve the phase field model, $\phi$ initialization is needed.
To apply local refinements, the model first needs to know $\phi$ initialization on a uniform coarse control mesh (Fig.~\ref{fig: truncated t-spline}A), refine elements at $\phi$ interface, and then obtain $\phi$ initialization on the locally refined mesh.
Beginning with a uniform coarse control mesh, we can calculate $\phi$ initialization based on the initial cell center and cell radius $r_0$.
Then, by selecting elements along $\phi$ interface (Fig.~\ref{fig: truncated t-spline}B), we can refine and enhance the effective mesh resolution at these critical areas of interest (Fig.~\ref{fig: truncated t-spline}C) while keeping the mesh coarse in areas where $\phi$ values are stable at either 0 or 1. 
This targeted refinement approach ensures efficient use of computational resources.
Because a valid T-mesh in truncated T-splines must be strongly balanced and free of face-face intersections to satisfy linear independence~\cite{wei2017truncated,buffa2010linear}, we need to resolve these face-face intersections in the automated local refinement process (Fig.~\ref{fig: truncated t-spline}D).
To address this issue, we apply bisections to 1-ring of elements surrounding the selected interface elements (Fig.~\ref{fig: truncated t-spline}E). 
This approach effectively resolves face-face intersections and allows the model to refine the mesh only where needed. 
Thus, we can obtain a locally-refined T-mesh for truncated T-splines tailored explicitly to the given $\phi$ value (Fig.~\ref{fig: truncated t-spline}F). 
% However, it should be noted that we constrain local refinement to a single level difference in neighboring elements due to the complexities involved with resolving intersections when extending to multiple levels.
% This choice is made primarily because addressing intersections across different levels of local refinements requires sophisticated and thorough applications of bisections. 
% Since the NDDs model demands automated truncated T-spline generation, repeated automatic local refinement throughout the simulation is necessary.
% Although applying bisection elements for multiple levels of local refinements is proven successful for steady-state analyses~\cite{wei2017truncated} and the concept is scientifically straightforward, it is not feasible in this case due to the large number of operations involved and its practical implementation is quite challenging.
Subsequently, $\phi$ values are interpolated and transferred onto this newly refined mesh using the KD tree-based interpolation method~\cite{bentley1975multidimensional},
%which we will discuss later in Section~\ref{sec: KD tree}, 
providing accurate initialization for the computation of the evolving neuron outgrowth $\phi$ (Fig.~\ref{fig: truncated t-spline}G). 
Starting with an existing control point, a search using the KD tree data structure finds the nearest new control point. 
If a matching control point is found, the value at the input control point is directly used.
%If the control point found is outside the previously defined domain boundaries, the algorithm handles them according to initialization logic.
%based on the type of variable being interpolated. 
%For most variables, a default value of ``0'' is assigned. 
%For orientation $\theta$, a random value between ``0'' and ``1'' is generated to introduce variability.
If the new control point is situated among the existing points, the algorithm interpolates depending on their positions and connectivity in the control mesh (Fig.~\ref{fig: truncated t-spline}G). 
This conditional handling ensures values are correctly passed to the locally refined mesh.
The KD tree approach effectively utilizes spatial indexing and positions, preserving the initialization accuracy while minimizing unnecessary computations.

\section{Dynamic domain expansion}
\label{sec: domain expansion}

During neuron growth, neurites initiate along soma boundaries and then extend outwards toward the domain boundary.
If using conventionally fixed domain size, this nature of the neuron growth process requires excessive DOFs in largely static areas where $\phi$ remains ``0'', away from the center. 
To address this issue, the NDDs model incorporates dynamic domain expansion to accommodate the complex, ever-expanding neurite elongation process, followed by a KD tree-based interpolation to ensure fast and accurate variable pass-through~\cite{bentley1975multidimensional}. 
This method reduces redundant DOFs associated with neurite extension, enabling the simulation of complex, high-resolution neuron outgrowth and enhancing both accuracy and computational efficiency.

% \subsection{Directional domain expansion}

\begin{figure}[h]
    \centering
    \includegraphics[width=\textwidth]{
        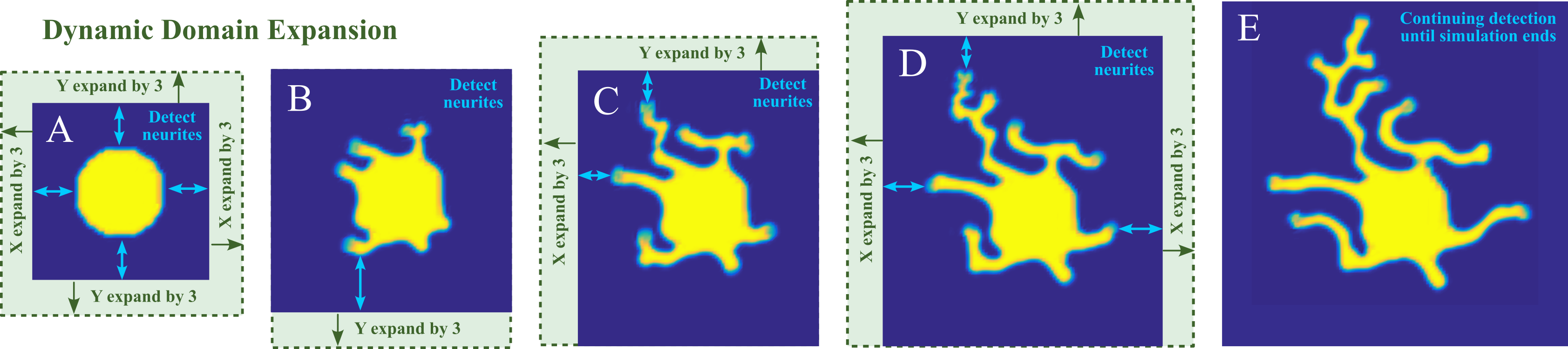
    }
    \caption{
        Sequential representation of directional domain expansion in neuron outgrowth simulation. 
        (A) The initial soma $\phi$ and checking for neurites near the domain boundary. 
        (B) Detected neurite near the bottom boundary and subsequent bottom directional domain expansion. 
        (C) As growth continues, neurites approach the top and left boundaries, and the domain is expanded along these directions. 
        (D) Neurites detected near the top, left, and right boundaries, and domain expansions follow. 
        (E) The expanded computational domain for neurite development. 
        The process showcasing the model can adapt to the evolving morphology of the neuron and minimize unnecessary computational costs.
    }
    \label{fig: dynamic domain expansion}
\end{figure}

The model incorporates a dynamic domain expansion algorithm (Algorithm \ref{Algorithm: dynamic_domain_expansion_algorithm}) to dynamically adjust the control mesh in response to neurite approaching the domain boundary (double-sided arrows in Fig.~\ref{fig: dynamic domain expansion}A-D).
The algorithm takes neuron growth $\phi$ and control mesh as the input and evaluates each element adjacent to the domain boundary, searching for any nonzero values of the $\phi$. 
If nonzero $\phi$ is detected near the boundary, the algorithm flags the corresponding edge for expansion using $expFlag$.
Once boundaries are flagged, the algorithm modifies the domain size based on the predetermined expansion size (default $3\times\Delta x$, where $\Delta x$ is coarse element size). 
It generates an enlarged control mesh (dashed zone in Figure \ref{fig: dynamic domain expansion}A-D) and translates the entire control mesh according to the specified direction of expansion to maintain the position of the neuron $\phi$. 
This directional domain expansion approach works in tandem with the truncated T-splines generation, discussed in Section~\ref{sec: truncated t-splines}, to ensure that the computational domain adapts to accommodate growing neurites with the least amount of elements, maintaining analysis accuracy and efficiency.
% \subsection{KD tree-based value interpolation for locally refined mesh}
% \label{sec: KD tree}
The transition to the expanded mesh requires an interpolation operation that passes values from the original control points to the new control points (Fig.~\ref{fig: truncated t-spline}G).
The interpolation process for a dynamically expanding mesh in the NDDs simulation leverages the KD tree algorithm to optimize spatial queries of control points $cpts$, significantly accelerating the search for the nearest $cpts$~\cite{bentley1975multidimensional}.
%Starting with new control points $cpts'$, the model searches for $cpts$ around it and uses values on $cpts$ to interpolate values on $cpts'$. 
Starting with new control points $cpts'$, each $cpts'$ undergoes a rounding process to identify whether it is located on expanded elements. 
This procedure significantly speeds up the algorithm by skipping elements that are not expanded, and reduces the total number of DOFs while preserving the analysis accuracy.

% A search follows using the KD tree to find the nearest existing control points $cpts$. 
% If a matching control point is found, the value from the input vector is directly assigned.
% For $cpts'$ outside the previously defined domain boundaries, the algorithm handles them according to initialization logic.
% %based on the type of variable being interpolated. 
% %For most variables, a default value of ``0'' is assigned. 
% %For orientation $\theta$, a random value between ``0'' and ``1'' is generated to introduce variability.
% For $cpts'$ among the existing points, the algorithm interpolates depending on their positions and connectivity in the control mesh (Fig.~\ref{fig: truncated t-spline}G). 
% This conditional handling ensures that values are correctly passed to the expanded mesh, maintaining the continuity and integrity of the simulation.
% The KD tree approach effectively utilizes spatial indexing and positions to optimize mesh interpolation, preserving the simulation resolution and accuracy while minimizing unnecessary computations.

\algrenewcommand\algorithmicrequire{\textbf{Procedure}}
\begin{algorithm}[h]
    \caption{Dynamic Domain Expansion (Figure~\ref{fig: dynamic domain expansion})}
    \textbf{Input}: Control points $cpts$, control mesh $Mesh$, neuron outgrowth $\phi$ \\
    \textbf{Output}: Expanded control points $cpts'$, Expanded control mesh $Mesh'$

    \begin{algorithmic}[1]
        \Require{Check for potential expansion direction based on $\phi$ value}
        \State Initialize expansion flag $expFlag$ for each element $e$ of the domain.
        \For{each element $e$ in $Mesh$ adjacent to the domain boundary}
            %\For{each node $(x, y)$ within clearance threshold $\delta$ from $e$}
                \If{$\phi(x, y) > 0$} 
                    \State $expFlag[e] = \text{true}$
                    \State \textbf{break} \Comment{Terminate search if expansion is necessary}
                \EndIf
            %\EndFor
        \EndFor
        \Require{Expanding domain based on expansion parameters}
        \If{any $expFlag[e]$ is true}  \Comment{Expand mesh where elements are flaggeed}
            \State Determine $dir_{exp}$ and $sz_{exp}$ based on flagged elements
            \State Create enlarged control mesh $Mesh'$ with new control points $cpts'$ 
            \State Translate $cpts'$ based on $dir_{exp}$ \Comment{Offset mesh to keep $\phi$ position consistant}
        \Else
            \State Set $cpts' = cpts$ and $Mesh' = Mesh$  \Comment{No expansion needed}
        \EndIf
        \State \Return $cpts'$ and $Mesh'$
    \end{algorithmic}
    \label{Algorithm: dynamic_domain_expansion_algorithm}
\end{algorithm}

\section{Experimental human iPSC-derived and rat hippocampal neuron cultures}
\label{sec: experimental culture}

In this section, we outline the protocols and details for culturing human induced pluripotent stem cells (iPSCs) and rat hippocampal neuron cultures.
Conducting experimental cultures to obtain neuron growth observations is necessary for computational modeling as it provides a reference to simulations and for comparison with realistic results.
This approach allows us to better understand the intricacy of neurite outgrowth, improving the ability of the NDDs model to reflect real-world biological behaviors in NDDs studies.
Human iPSC-derived neurons are selected to better correlate between the NDDs model and actual human neuron growth. 
Building upon our previous studies of healthy rat hippocampal neurons~\cite{qian2023biomimetic,liao2023semi}, we now extend our focus to unhealthy rat hippocampal neurons to deepen our understanding of NDDs.
In future work, we intend to include unhealthy human iPSC-derived neurons currently being developed at the Mayo Clinic to broaden our investigation of underlying disorder mechanisms.

\textbf{Human iPSC-derived neuron culture.}
% \textcolor{blue}{Rephrased and shortened version, asked Genesis to rephrase, will update if they finish one version.}
Human iPSCs, derived from a healthy 83-year-old female (MC0192), were cultured in TeSR-E7 complete medium on Matrigel-coated plates, with karyotyping conducted by the Mayo Clinic Genomics Core Facility~\cite{zhao2017apoe}. 
The iPSCs were differentiated into neural precursor cells (NPCs) using the STEMdiff\textsuperscript{\texttrademark} SMADi Neural Induction Kit as previously described~\cite{kawatani2023abca7}. 
After initial culturing in 24-well AggreWell\textsuperscript{\texttrademark} 800 plates, embryoid bodies were transferred to Matrigel-coated 6-well plates on day 8 to promote neural rosette formation for another 5 days. 
These NPCs were expanded, supplemented with ROCK inhibitor Y27632 on day 15, and eventually cryopreserved in Neural Progenitor Freezing Medium. 
NPCs were seeded on Poly-L-ornithine/Laminin-coated plates for neuronal differentiation in BrainPhys Neuronal Medium. 
Immunocytochemistry involved fixing the cells, staining them with primary antibodies against $\beta$III-tubulin and secondary Alexa Fluor 488- or 568-conjugated secondary antibodies, and counterstaining nuclei with DAPI. 
Imaging was performed using a Keyence BZ-X800 fluorescence microscope.
We refer readers to~\cite{zhao2017apoe,kawatani2023abca7} for detailed documentation of experimental protocols.
The images of experimental human iPSC-derived neuron culture are used to compare with healthy neuron growth simulations to better analyze neurite outgrowth behaviors.

\textbf{Rat hippocampal neuron cultures.}
For rat hippocampal neuron culture results used in this paper, images are drawn from failed cultures during protocol optimization for our previously published work creating a dataset and model of rat hippocampal neuron growth~\cite{qian2023biomimetic,liao_quantitative_2022} and from cultures in which cells experienced mechanical damage.

For culture condition optimization examples, 
% For the spontaneous degeneration examples,
cryopreserved embryonic day 18 neurons (A36513, Gibco, USA) were thawed and grown in dishes coated with poly-D-lysine (PDL) (P6407, Sigma-Aldrich, USA) following the manufacturer's guidelines~\cite{2018B-27System} with slight modification to optimize conditions for low density culture. Primary neurons are notoriously challenging to culture~\cite{gordon2021general}.
% (ADD REF). 
It is not uncommon when establishing new neural cultures to optimize media formulation, media change frequency, and the fraction of media replaced during media changes. During this optimization process several cultures underwent cell death with neurite breaking and fragmentation. 
The seeding density for all samples was either 20,000 or 100,000 cells per square centimeter in Neurobasal Plus medium (A3582901, Gibco, USA) supplemented with 2\% B-27 Plus (A3582801, Gibco, USA). 
Cultures were maintained at 37$^{\circ}$C with 5\% $CO_2$. 
Imaging was carried out using an Echo Revolve Microscope (Echo Revolve $\vert$ R4, inverted, BICO, USA) with a 12-megapixel color camera at 20X and 40X magnifications. The images included here for model comparison were all taken 3-4 days after plating.

For samples in which neurons experienced mechanical damage, wells were coated first with PNIPAM following manufacturer protocols and subsequently coated with a PDL-ECM gel mixture (PECM) composed of Geltrex ECM Gel (Thermo Fisher, A1413301) diluted 1:100 in DMEM and mixed with 10 $\mu$g/mL PDL~\cite{yamamoto2016unidirectional,vogt2003micropatterned}
% (REFS) 
for 1 hour at room temperature. The coating solution was then removed and the coated dishes were left uncovered in the laminar biosaftey cabinet for 2 hours. Cryopreserved embryonic day 18 neurons (A36513, Gibco, USA) were then seeded at a density of 20,000 cells/cm$^2$ and cultured for 2 days at 37$^{\circ}$C in 5\% $CO_2$. After the initial culture period cells were lifted from the plate using a gelatin plunger transfer technique~\cite{haraguchi2012fabrication}.
% (REF). 
Briefly, a gelatin gel plunger was gently placed in contact with the cells and the cells were released from the plate by incubation at 20$^{\circ}$C. The gelatin plunger was then removed and placed into a separate PDL coated well. This new plate was briefly incubated at 20$^{\circ}$C followed by incubation at 37$^{\circ}$C to liquefy the gelatin. All wells were gently washed with Hank's balanced salt solution to remove gelatin and imaged.

It should be noted that these conditions were not intended to replicate any specific developmental disorder. They are included here for qualitative comparison to demonstrate the models ability to create degenerating and damaged morphologies.

\section{Numerical results and validation with experiments}
\label{sec: results}

In this section, we first explain the implementation specifics of the proposed IGA phase field NDDs model. 
Then, we showcase simulation results of healthy and unhealthy neurons with several parameter studies designed to investigate NDDs. 
Finally, we compare our simulation outcomes with experimental data, focusing on cases with external cue-guided mechanisms. 
This comparative analysis aims to evaluate the effectiveness and accuracy of our NDDs model in simulating biomimetic neurite behaviors and disorders.

We implemented the IGA NDDs model in C++, leveraging the extensive capabilities and scalabilities of the PETSc library~\cite{petsc-user-ref}. 
PETSc as the computational backbone offers a significant computational efficiency advantage over MATLAB-based implementations~\cite{qian_modeling_2022}.
This is mainly because the PETSc library is written in compiled language and tailored for high-performance computing environments, while MATLAB is a scripting language that faces limitations due to its inherently interpreted mode of execution, making it less efficient for tasks requiring intensive computation.
Through the utilization of the MPI for parallel processing, our model facilitates efficient distribution of computations across multiple threads~\cite{petscsf2022,gropp1996high}, thereby cutting down execution times significantly when compared with MATLAB analysis. 
Although MPI leverages the immense computational power enabled by supercomputers, communication time among processors becomes a bottleneck. 
%Based on the locally-refined truncated T-spline, m
Mesh partitioning using METIS~\cite{karypis1998fast} ensures optimal load balancing for the computationally intensive simulations. 
This is crucial for the scalability and efficiency of the model, particularly when simulating complex neurite structures on locally refined truncated T-splines with a large number of DOFs because mesh alterations during execution often lead to unbalanced loads. 
Unbalanced loads across threads could lead to unnecessary slowdown and waiting during inter-threads/node communications. 
We ran simulations on the Bridges-2 supercomputer at Pittsburgh Supercomputing Center~\cite{ecss, xsede} using 128-thread regular memory nodes.

% white background cells are added to ensure text vertical alignment
% \vspace{-0.5cm}
\begin{table}[ht]
\caption{Parameters used in the NDDs model.}
\vspace{-0.7cm}
\label{Table: NDDs parameter}
\begin{center}
\resizebox{\columnwidth}{!}{
\begin{tabular}{c l c|c l c}
\toprule
\textbf{Parameter} & \textbf{Description} & \textbf{Value} & \textbf{Parameter} & \textbf{Description} & \textbf{Value}\\
\hline
\cellcolor{grayshade} $c_{opti}$ & \cellcolor{grayshade} Optimal neurotrophin concentration & \cellcolor{grayshade} [0, 1] & $H$ & Orientation constant coefficient & $0.007$ \\
\cellcolor{grayshade} $D_c$ & \cellcolor{grayshade} Neurotrophin diffusion coefficient & \cellcolor{grayshade} [0, 6] & $\delta_t$ & Tubulin diffusion rate & $4$ ($\mu m^2/h$) \\
\cellcolor{grayshade} $k_{p75}$ & \cellcolor{grayshade} Neurotrophin binding rate & \cellcolor{grayshade} [0, 3] & $\alpha_t$ & Tubulin active transport rate & $0.001$ ($\mu m/h$) \\
\cellcolor{grayshade} $k_2$ & \cellcolor{grayshade} Neurotrophin degradation rate & \cellcolor{grayshade} [0, 3] & $\beta_t$ & Tubulin decay coefficient & $0.001$ ($1/h$) \\
\cellcolor{grayshade} $\gamma$ & \cellcolor{grayshade} Phase field interface driving force constant & \cellcolor{grayshade} [0, 10] & $\epsilon_0$ & Tubulin production coefficient & $15$ \\
\cellcolor{grayshade} $K$ & \cellcolor{grayshade} Dimensionless latent neurotrophin & \cellcolor{grayshade} [0, 2] & $r_0$ & Initial cell radius & $15\triangle x$ \\
\cellcolor{white} $M_{\phi}$ & \cellcolor{white} Mobility coefficient & \cellcolor{white} $60$ & $r_g$ & Tubulin assembly rate & $5$  \\
\cellcolor{white} $\bar{a}$ & \cellcolor{white} Surface energy scaling constant & \cellcolor{white} 0.04 & $s_g$ & Tubulin disassembly rate & $0.1$ \\
\cellcolor{white} $\xi$ & \cellcolor{white} Anisotropy strength & \cellcolor{white} 0.2 & $\frac{\alpha}{\pi}$ & Scaling coefficient & $0.2865$ \\
\cellcolor{white} $k$ & \cellcolor{white} Anisotropy mode & \cellcolor{white} 6 & $\theta$ & Neurite growth orientation angle & $[0,1]$ \\
\bottomrule
\multicolumn{6}{l}{Note: \colorbox{grayshade}{Variables shaded in gray} contribute to NDDs morphological transformation. For $c_{opti}$, higher levels cause excessive branching} \\
\multicolumn{6}{l}{and lower values lead to retractions. 
For $D_c$, high levels initiate neurite formation and low levels lead to atrophy.
$k_{p75}$ increases neurite} \\
\multicolumn{6}{l}{thickness as values rise. 
$k_2$ increases neurite thickness as values rise. 
$\gamma$ maintains a constant effect on interface stability and scales the} \\
\multicolumn{6}{l}{effect of $c_{neur}$. Higher $K$ values enhance neurite thickness and branching. Parameters require initializations are shown with their default } \\
\multicolumn{6}{l}{value. Dimensionless parameters are listed with default values without units. $\Delta x$ is the coarse element size} \\
\end{tabular}}
\end{center}
\end{table}

% Low levels of $c_{opti}$ will lead to low driving force value, which retracts the $\phi$ interface.
% Limited neurotrophin diffusion will lead to insufficient neurotrophin concentration to support neuron survival.
For clarity, we provide a detailed list of variables for NDDs and healthy neurons in Table~\ref{Table: NDDs parameter}. 
The domain $\phi$ is initialized with a central filled circle with radius $r_0$, representing the cell.
The initial values of $\theta$ are randomized to between $[0, 1]$ across the domain with $T$ set to 0 at the beginning.
Due to the significant variability in biophysics and growth behaviors across different types of neurons, the parameter values used in the phase field NDDs model were sourced from established literature~\cite{takaki_phase_field_2015, nella2022bridging, diehl2016efficient} and fine-tuned empirically to capture realistic neuron growth. 
These parameters can be adjusted to better simulate biomimetic growth of specific neuron types
~\cite{qian2023biomimetic}.
With these parameter settings, we first simulate healthy neuron growth and analyze single- and multiple-neuron growth scenarios (Section~\ref{sec: healthy simulation}).
Then, we conduct a simulation of abnormal neurite morphological transformation using the NDDs model with three case studies that each focuses on a specific parameter (Section~\ref{sec: disorder study}).
Finally, we compare simulation results with experimental observations to evaluate our NDDs model (Section~\ref{sec: comparison study}).

\subsection{Healthy neuron growth simulation}
\label{sec: healthy simulation}

\begin{figure}[!h]
    \centering
    \includegraphics[width=\textwidth]{
        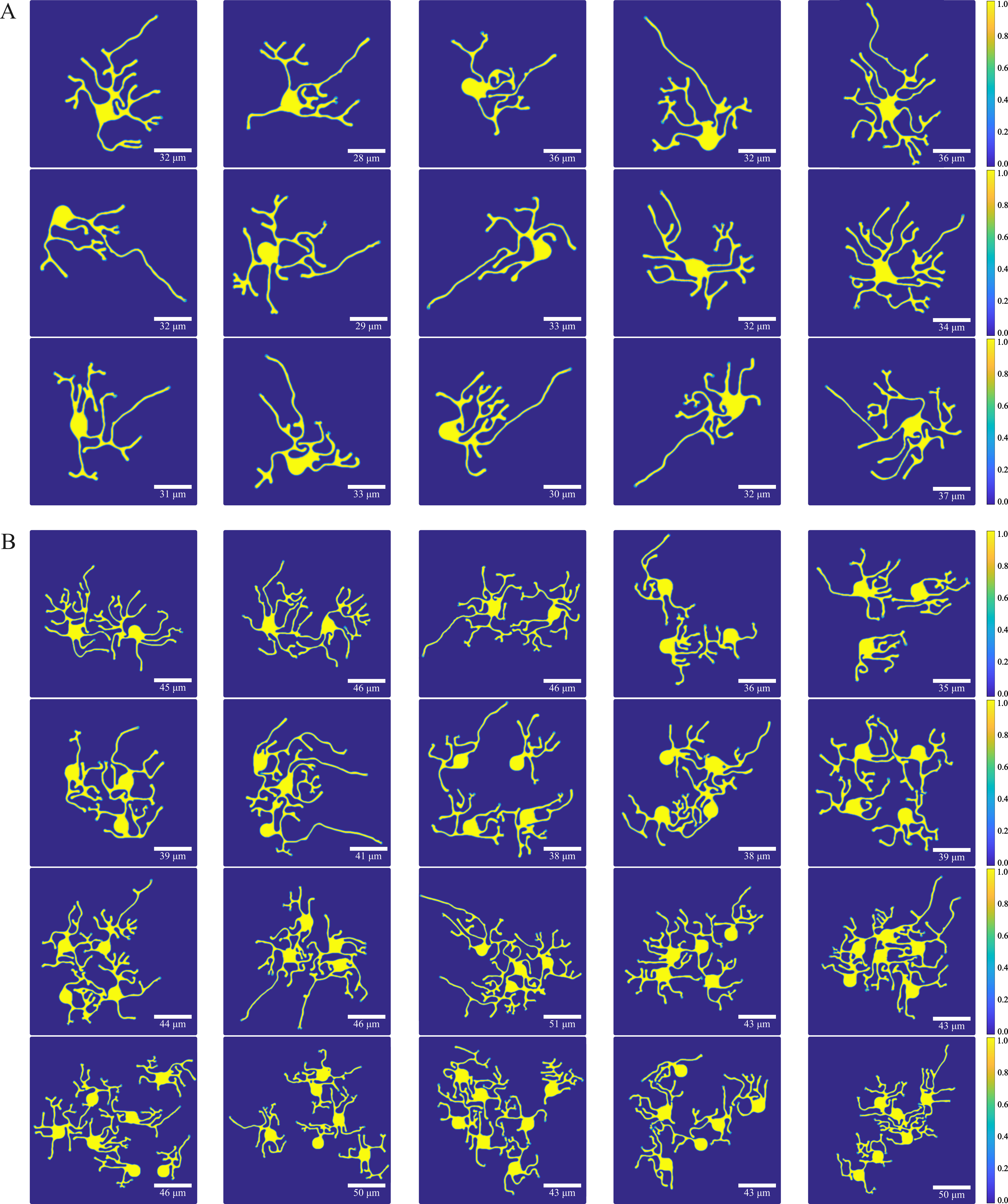
    }
    \caption{
        Healthy neuron growth simulations.
        (A) Single neuron growth with many neurite morphologies. 
        (B) Multiple-neuron growth simulations with neurite interactions. For multi-neuron cases, the initial soma placements are randomized in the domain.
    }
    \label{fig: health neuron}
\end{figure} 

Our NDDs model supports simulating healthy neuron growth. 
We set $c_{opti}$ to 1 to ensure sufficient concentration level to drive interface outwards expansion, $D_c$ to 6 to ensure sufficient neurotrophin diffusion, and $k_{p75}, k_2 c_{neur}$ to 0 to eliminate degradation effects.
The rest parameters for healthy neuron growth are set following Table~\ref{Table: NDDs parameter}.
Simulations are done with both single- and multiple-neuron configurations up to 350,000 iterations, and results near the final iteration that best illustrate the growth behaviors are shown (Fig.~\ref{fig: health neuron}).
The single-neuron scenarios aim to study individual neuron outgrowth behaviors, while the multiple-neuron setups focus on understanding neurite interactions that contribute to the formation of complex neurite networks. 

In our results, the neuron, represented by $\phi$, is shown in yellow, and the surrounding medium is shown in blue.
For single neuron cases, the results (Fig.~\ref{fig: health neuron}A) show that the model captures a broad spectrum of neurite outgrowth behaviors, from axon elongations to complex branching. 
%These results display less branching and capture more natural neurite progression compared to the existing model~\cite{qian_modeling_2022}. 
Concurrently, our multiple-neuron simulations investigate interactions within complex neurite network formation, varying from 2-7 neurons (Fig.~\ref{fig: health neuron}B). 
Starting from 2-neuron cases, we gradually increase complexity to configurations with up to 7 randomly placed neurons. 
These multi-neuron simulations showcase individual neurite growth patterns and the intricate neurite interactions with responses of other neurons in proximity. 
With the detailed dynamics of $c_{tubu}$ and $c_{neur}$, these biomimetic single and multi-neuron simulation results showcase the potential of our model for in-depth neurodevelopmental research and provide a possible computational foundation for exploring NDDs.

\subsection{Neurodevelopmental disorders}
\label{sec: disorder study}

\begin{figure}[!h]
    \centering
    \includegraphics[width=\textwidth]{
        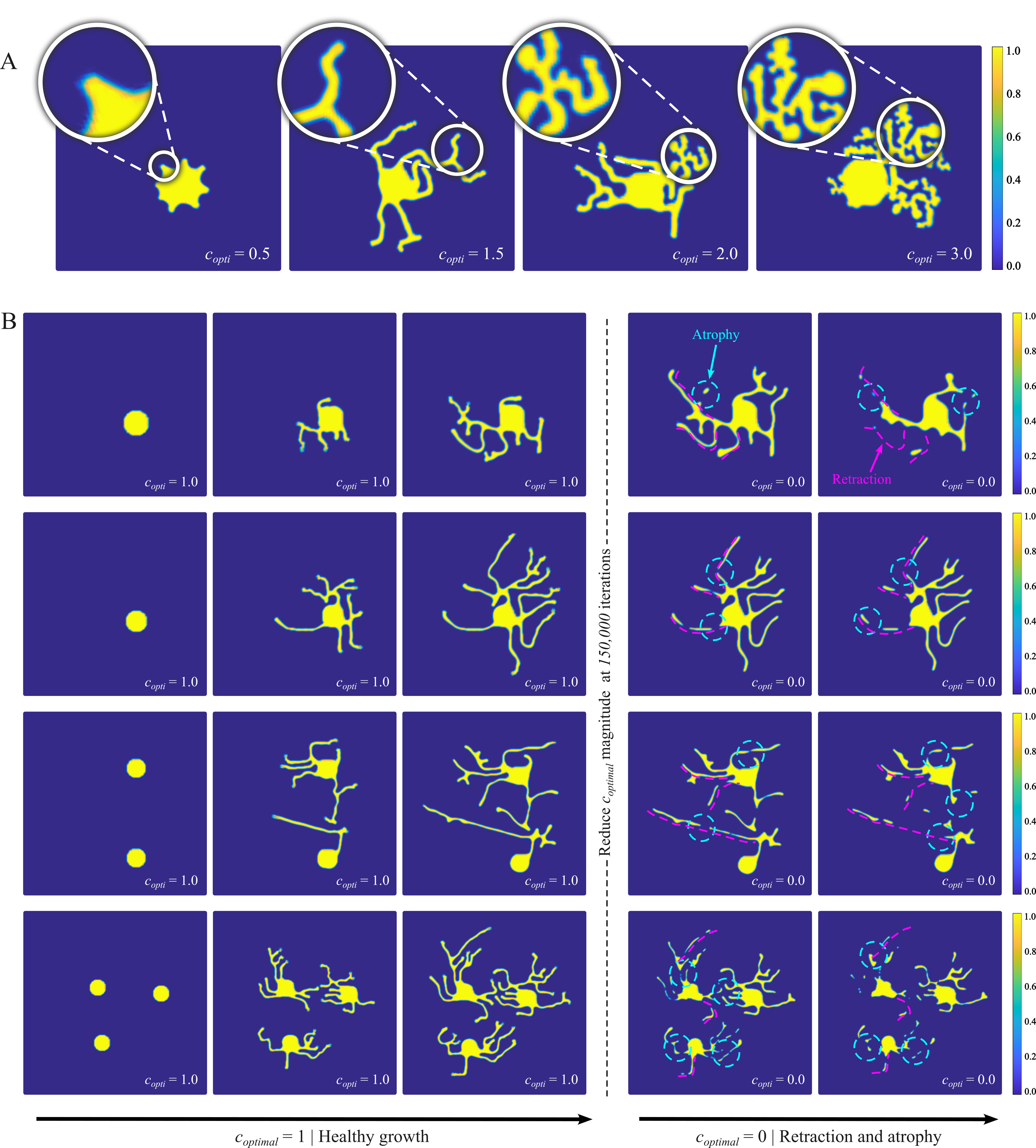
    }
    \caption{
        Impact of $c_{opti}$ variations on neurite morphology and development.
        (A) Variations of neurite growth behaviors as influenced by $c_{opti}$ values ranging from 0.5 to 3.0 with zoomed-in views of the intricate neurite structures at around 40,000 iterations, illustrating the progressive neurite density and branching complexity changes. 
        (B) Demonstration of the dynamic retraction behaviors in the simulated neuron growth process with an initial $c_{opti}$ value of 1, which is subsequently reduced to 0 at 150,000 iterations during the simulation to mimic the effect of increasing $c_{neur}$ magnitude and its inverse relationship on neuron survival. 
        Atrophies are marked with cyan dashed circles, and retractions are traced with magenta dashed lines.
        The $c_{opti}$ reduction simulates neurite retraction, showcasing the potential impacts of decreasing neurotrophic support on neurite morphology and structural integrity over time.
        }
    \label{fig: Disorder Copt}
\end{figure}

\begin{figure}[h]
    \centering
    \includegraphics[width=\textwidth]{
        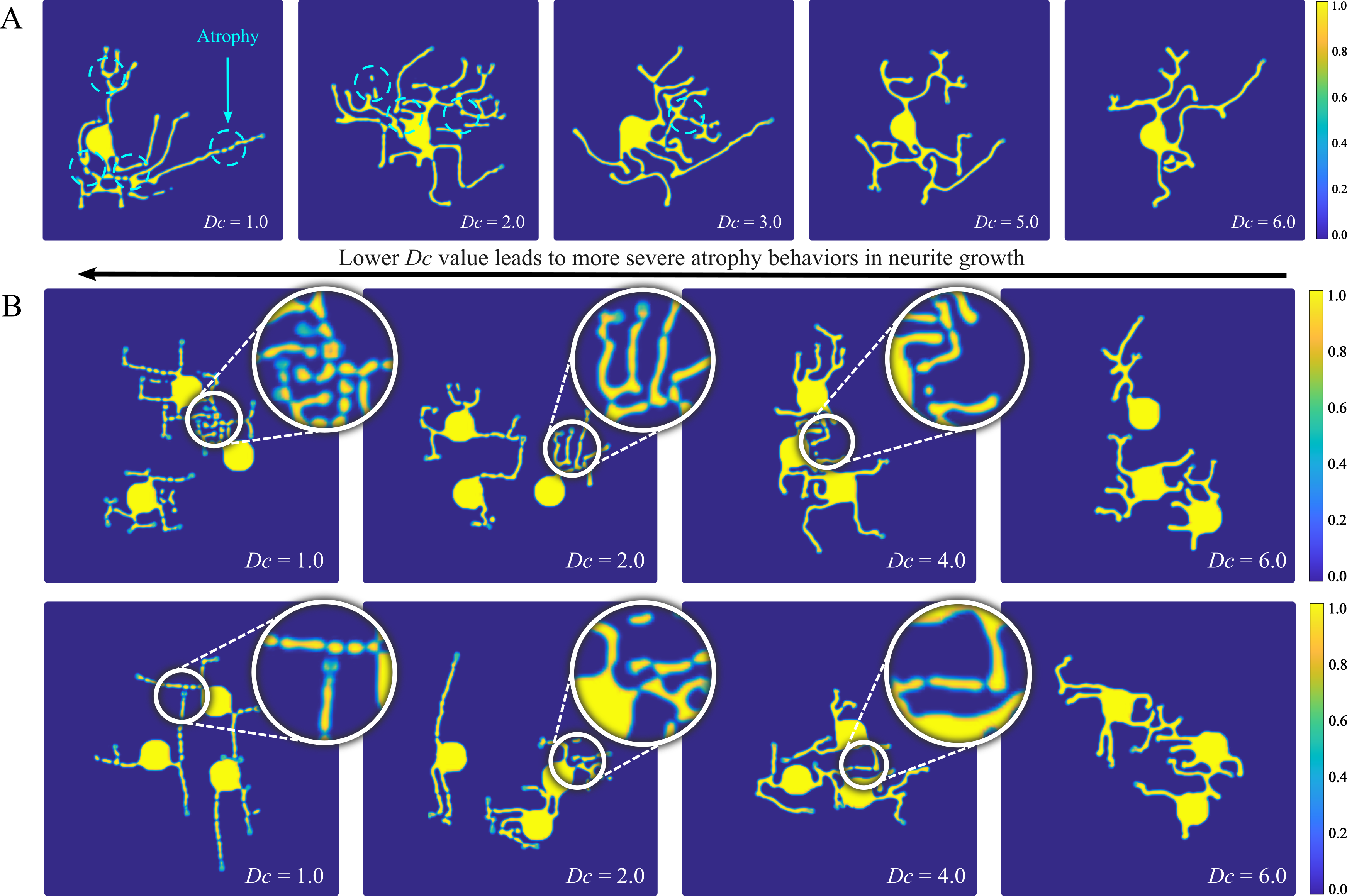
    }
    \caption{
        Impact of neurotrophin diffusion rate $D_c$ variations on neurite outgrowth. 
        (A) Single- and (B) multiple-neuron NDDs simulations with different $D_c$ values.
        It highlights how low values of $D_c$, corresponding to inadequate diffusion, can lead to neurite atrophy. 
        As the $D_c$ value increases from $1.0$ to $6.0$, neurite continuity improves noticeably, showcasing the critical role of diffusion rates in supporting and maintaining neurite structures.
        Atrophies are marked in (A) with cyan dashed circles.
        Zoomed-in views in (B) provide a detailed view of atrophy behaviors.
    }
    \label{fig: Disorder Dc}
\end{figure}

\begin{figure}[h]
    \centering
    \includegraphics[width=\textwidth]{
        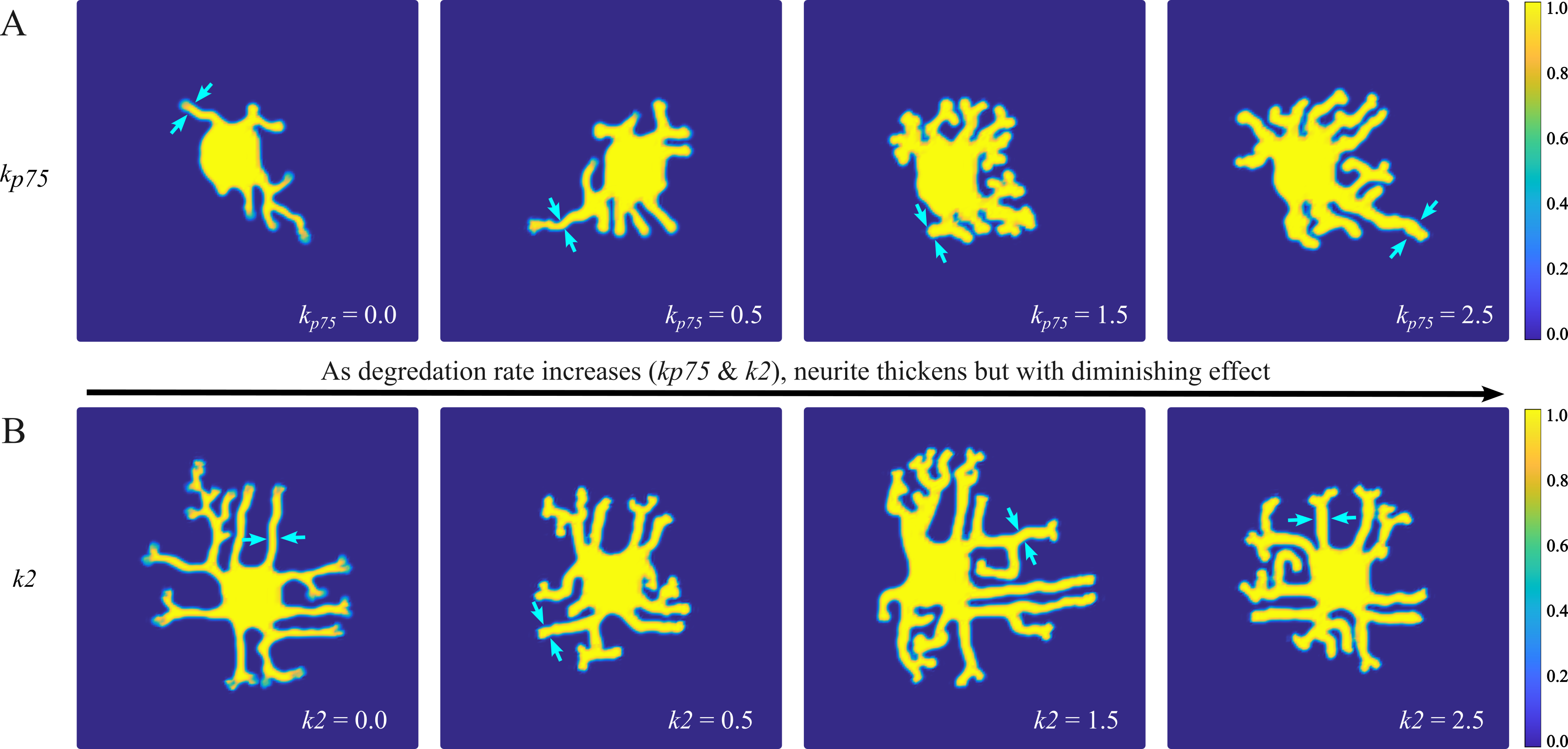
    }
    \caption{
        Impact of variations in degradation rates $k_{p75}$ and $k_2$ on neurite morphology.
        (A) Neurite growth results with varying $k_{p75}$ values.
        (B) Neurite growth results with varying $k_2$ values.
        Both variables predominantly affect neurite thickness (marked with cyan arrows), subtly impacting the physical structure while maintaining the complexity of the neurite structures unchanged. 
        Increasing values of these variables result in progressively thicker neurites but with diminishing effects.
    }
    \label{fig: Disorder k}
\end{figure}

Extending from healthy neuron growth simulations, we conduct a comprehensive study (Figs.~\ref{fig: Disorder Copt}-\ref{fig: Disorder k}) to investigate the functional roles of different parameters and factors affecting the onset of NDDs, with a focus on critical parameters $c_{opti}$, $D_c$, $k_{p75}$, and $k_2$ as detailed in Table~\ref{Table: NDDs parameter}. 
In this section, we categorize the parameters into three studies based on their effects on neurite morphological transformations: 
\begin{enumerate}
  \item Optimal neurotrophin concentration $c_{opti}$. This parameter models the inverse relationship between neurotrophin concentration level and neuron survival and is analyzed for its impact mostly on neurite retraction;
  \item Neurotrophin diffusion rate $D_c$. This parameter is responsible for diffusing the concentration necessary for neurite outgrowths and predominantly triggers neurite atrophy; and
  \item Degradation rates $k_{p75}$ and $k_2$. These parameters are investigated for their effects on neurite thickness. Very thin neurites lead to atrophy.
\end{enumerate}
%$c_{opti}$, which is analyzed for its impact mostly on neurite retraction; (2) $D_c$, which predominantly triggers neurite atrophy; and (3) degradation rates $k_{p75}$ and $k_2$, which are investigated for their influence on the thickness of neurite growth.
For each study, the parameter being analyzed was adjusted based on the specified ranges listed in Table~\ref{Table: NDDs parameter}. 
All other parameters were set according to their default values listed in the same table.
By adjusting these parameters in our NDD simulations, we can not only enhance our understanding of the pathophysiological mechanisms of NDDs but also examine the capability of our model to capture the intricate dynamics, which is critical for future application of the NDDs model in developing targeted interventions and advancing neurodevelopmental research.

\textbf{Optimal neurotrophin concentration $c_{opti}$.}
%In this study, we focus on the effect of optimal neurotrophin concentration $c_{opti}$.
By adjusting the value of $c_{opti}$ while keeping other parameters consistent with Table~\ref{Table: NDDs parameter}, we can better understand its impact on neurite morphological transformations in NDDs.
% Fig.~\ref{fig: Disorder Copt} showcases the impact of the optimal neurotrophin concentration, $c_{opti}$, on the morphogenesis of neurons. 
We first analyze neurite growth behaviors under different $c_{opti}$ values ranging from $0.5$ to $3.0$ at around $40,000$ iterations (Fig.~\ref{fig: Disorder Copt}A).
As $c_{opti}$ increases from $0.5$ to $1.5$, there is a noticeable improvement in neurite density and complexity, promoting more extended growth patterns. 
This process indicates a positive response of neurites to optimal neurotrophin levels, matching our previous expectations of enhanced growth behaviors. 
As $c_{opti}$ progresses from $1.5$ to $3.0$, there is a noticeable shift towards highly branched and extensively multidirectional neurite structures.
This range of growth patterns, from minimal to excessive branching, shows a critical threshold level of neurotrophins necessary for neurodevelopmental processes.
Notably, lower values of $c_{opti}$ lead to suppressed neurite outgrowth, validating that there is an inverse relationship between $c_{neur}$ and neuron survival~\cite{piontek1999neurotrophins}, as well as a critical balance of $c_{neur}$ for maintaining healthy neurite development. 
We then explore the dynamic effects of varying $c_{opti}$ in our NDDs model to understand its impact on neurodevelopmental processes.
For simulation cases in Fig.~\ref{fig: Disorder Copt}B, we initialize $c_{opti}$ as 1 to simulate healthy neurite outgrowth and then reduce $c_{opti}$ to $0$ at $150,000$ iterations across cases involving different numbers of neurons. 
The first two rows in Fig.~\ref{fig: Disorder Copt}B showcase the growth behavior of individual neurons, showing a straightforward process of neurite retraction as $c_{opti}$ decreases, highlighting the sensitivity of single neurons to $c_{neur}$ and $c_{opti}$ levels. 
The third and fourth rows extend this analysis to multi-neuron configurations with two and three neurons to investigate neurite retraction and interactions among multiple neurons. 
These simulation results highlight the intricate influence of $c_{opti}$ levels on the simulated morphology transformation of neurites and the potential implications to retraction and atrophy simulations, vital for informing the future development of effective therapeutic strategies targeting NDDs.

\textbf{Neurotrophin diffusion rate $D_c$.}
In this study, we focus on the neurotrophin diffusion rate $D_c$ that is responsible for controlling the diffusion of $c_{neur}$ to support $\phi$ interface balance.
We analyze the effect of $D_c$ with varying magnitude from $1.0$ to $6.0$ on single- and multiple-neuron cases, and results at around $100,000$ iterations that best exhibit atrophy are shown in Fig.~\ref{fig: Disorder Dc}.
% Fig.~\ref{fig: Disorder Dc} illustrates the pivotal role the neurotrophin diffusion rate, $D_c$, has on the neurodevelopmental process and its strong correlations to neurite atrophy. 
% Fig.~\ref{fig: Disorder Dc}A showcases single neuron simulation results using $D_c$ values ranging from $1.0$ to $6.0$.
The results show severe atrophy when the $D_c$ value is below $2.0$. 
As the $D_c$ value approaches $3.0$, atrophy becomes moderate.
This atrophy behavior indicates that lower values of $D_c$, particularly at 1.0, severely restrict the diffusion necessary for generating a neurotrophin gradient field for Eqn.~\ref{eqn: driving force equation} that stabilizes the phase field interface during neurite outgrowth. 
This insufficient gradient compromises neurite integrity and leads to the degeneration or atrophy of neurite structures due to inadequate neurotrophic support. 
As a result, as neurite tips continue growing out, there is not enough concentration to support and maintain the grown neurite structures, leading to neurite atrophy.
On the other hand, as $D_c$ increases, particularly beyond $5.0$, a noticeable improvement in neurite continuity is observed in both single- (Fig.~\ref{fig: Disorder Dc}A) and multiple-neuron cases (Fig.~\ref{fig: Disorder Dc}B), markedly reducing the incidence of neurite atrophy. 
This highlights the necessity of maintaining a diffusion rate $D_c$ to ensure a consistent concentration distribution surrounding the simulated growing neurite, thereby supporting the uninterrupted neurodevelopmental process during the simulation and providing insights for potential NDDs progression mitigation.

\textbf{Degradation rates $k_{p75}$ and $k_2$.}
%In this study, we focus on the degradation rates $k_{p75}$ and $k_2$.
We initialize the simulation with $k_{p75}$ and $k_2$ values ranging from $0.0$ to $2.5$, and results at around $30,000$ iterations are shown in Fig.~\ref{fig: Disorder k}.
% Fig.~\ref{fig: Disorder k} presents a detailed examination of the influence of degradation rate parameters $k_{p75}$ and $k_2$ on the morphological evolution of neurons simulated using our computational NDDs model. 
The parameter $k_{p75}$, when adjusted from $0.0$ to $1.5$, enhances the complexity of neurite branching, as well as from thin neurites to thicker neurites (Fig.~\ref{fig: Disorder k}A). 
This progression showcases the role of $k_{p75}$ in modulating neurotrophin generation and maintaining $c_{neur}$ levels, which are vital for the healthy development of neurite structures. 
As increasing $k_{p75}$ value increases beyond $1.5$ to $2.5$, the results showcase a plateau in morphological complexity changes, indicating a point where further increases in $k_{p75}$ yield diminishing effects on morphological changes.
Similarly, the impact of $k_2$ on neurite morphology is demonstrated through its influence on increasing neurite thickness (Fig.~\ref{fig: Disorder k}B), enhancing the structures of neurites but without subtle changes to the overall branching patterns and the complexity of the neurite structures. 
As the $k_2$ value rises, its effects also gradually diminish, highlighting diminishing effects similar to $k_{p75}$. 
These results provide an interesting perspective on the degradation of neurotrophin in the pathophysiology of NDDs, which will contribute to the development of targeted therapeutic planning.

\vspace{2mm}
These detailed parameter studies utilizing our NDDs model demonstrate NDDs morphological transformations to specific simulation parameters. 
The impact of $c_{opti}$ is most dominantly on neurite retraction and branching, the influence of $D_c$ leads to neurite atrophy, and the degradation rates $k_{p75}$ and $k_2$ affect the thickness of neurite growth. 
By categorically investigating these parameters, we have identified crucial factors that predominantly influence the progression of NDDs in our computational model.
Our analysis provides crucial insights into how variations in these key parameters can trigger or exacerbate the symptoms associated with NDDs. 
This knowledge is vital in guiding ongoing research toward understanding the pathophysiological mechanisms underlying these disorders. 
In the long term, this model could enable more targeted therapeutic planning by focusing on the most influential factors of neuron development and treatment strategy.
% This approach will enhance our understanding of potential NDDs triggering mechanisms and broaden the potential reach for effective treatment strategies by incorporating the effect of parameters with biophysical processes in simulating realistic neurodevelopmental progress.

\subsection{Qualitative validation with experiments}
\label{sec: comparison study}

\begin{figure}[h]
    \centering
    \includegraphics[width=\textwidth]{
        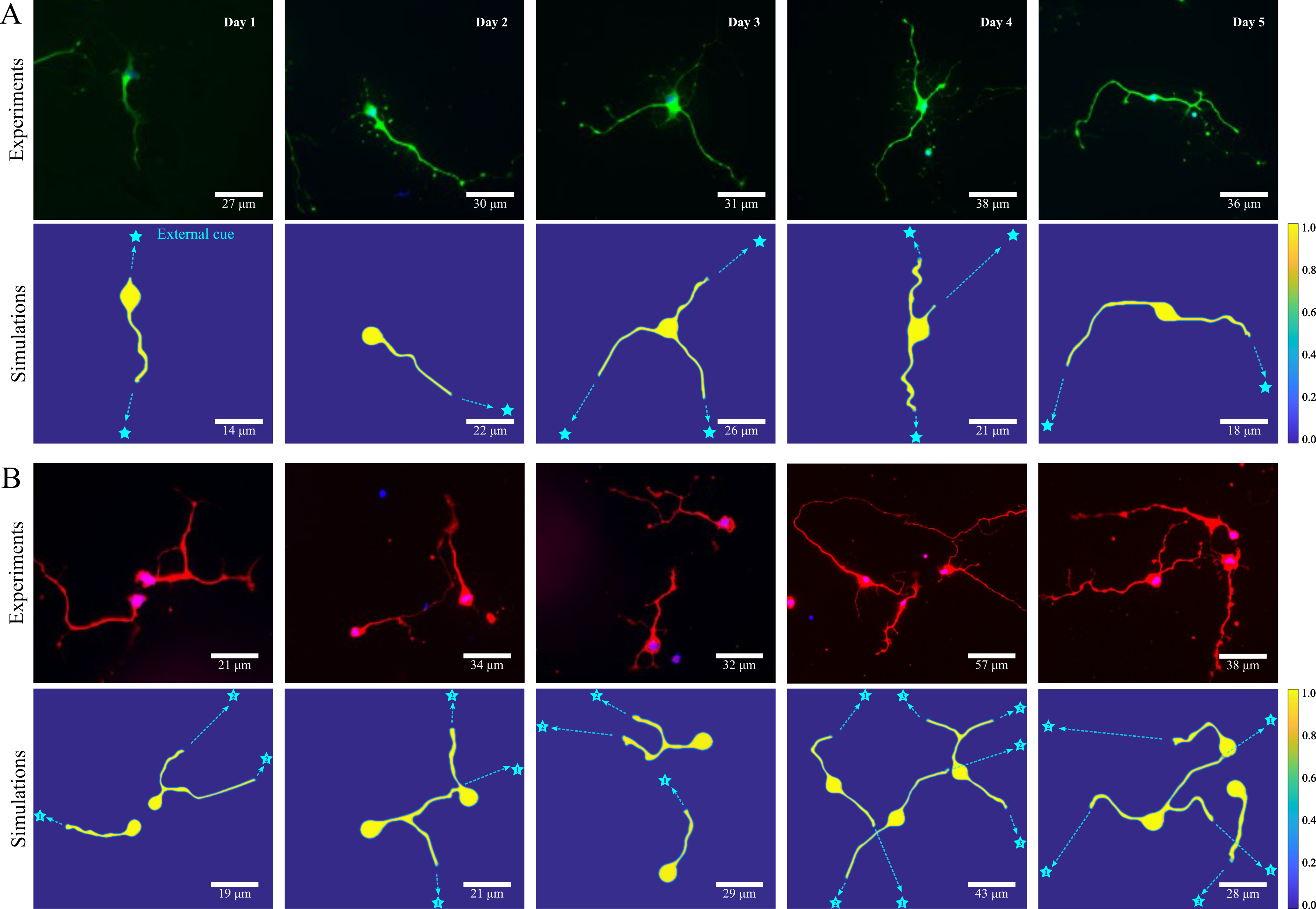
    }
    \caption{
        Comparison of simulated neuron growth patterns and experimental cultures of human iPSC-derived neurons. 
        (A) Comparisons of single neuron cases up to day 5.
        (B) Comparisons of multiple-neuron cases involving two and three neurons. 
        External cue placements are marked with cyan stars.
        These comparisons provide a visual qualitative validation of NDDs model to replicate key aspects of neurite morphology and developmental processes seen \textit{in vitro}.
    }
    \label{fig: Exp comparison results - healthy}
\end{figure}

\begin{figure}[h]
    \centering
    \includegraphics[width=\textwidth]{
        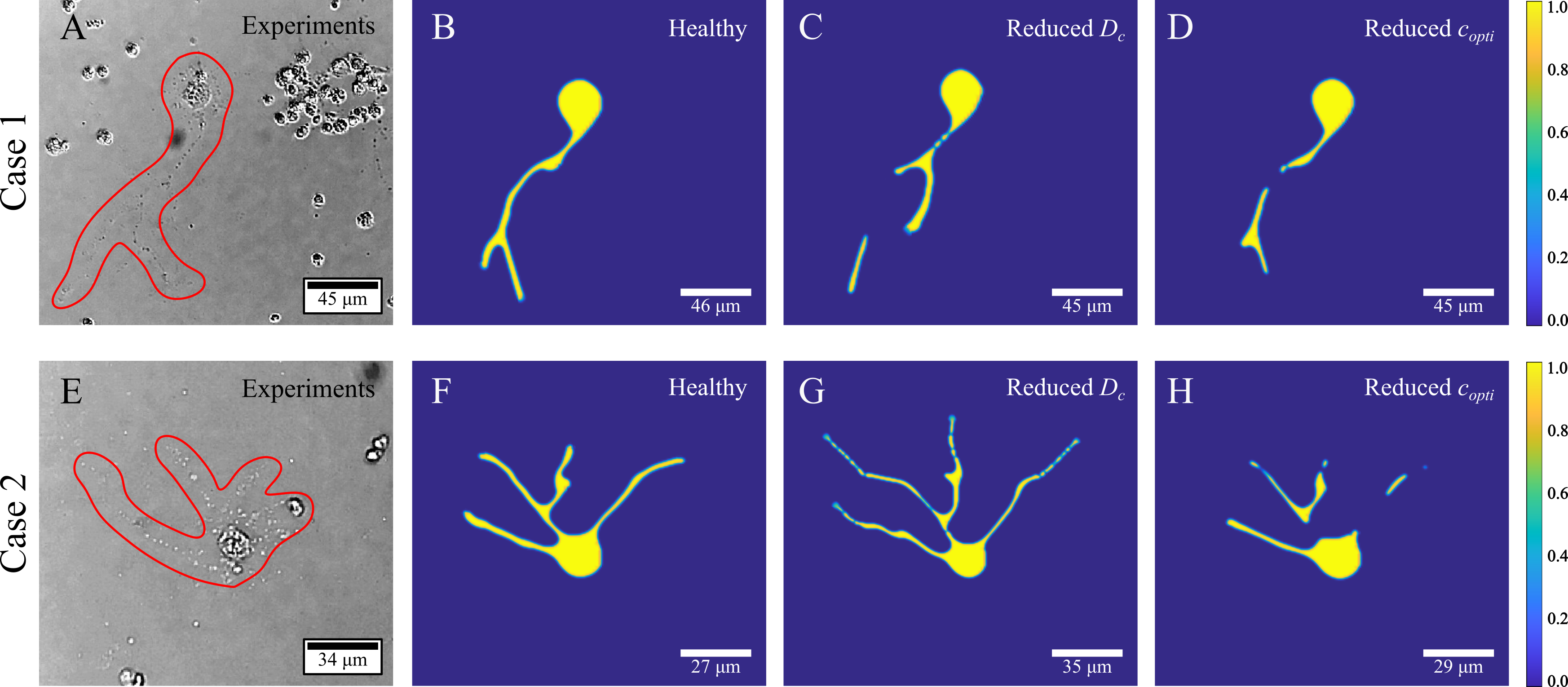
    }
    \caption{
    Comparison of experimental neuron cultures undergoing neurite degeneration and simulation results.
    (A\&E) Experimental images of neurons exhibiting neurite breaking and fragmentation
    % and degeneration 
    are highlighted in red outlines.
    (B\&F) Simulated healthy neurons using the NDDs model.
    (C\&G) Simulated neuron morphology with disorders induced by reducing the diffusion rate $D_c$ from 6 to 1.
    (D\&H) Simulated neuron morphology with disorders induced by lowering the optimal concentration $c_{opti}$ from 1 to 0.
    }
    \label{fig: Disorder singleCaseStudy}
\end{figure}

\begin{figure}[h]
    \centering
    \includegraphics[width=\textwidth]{
        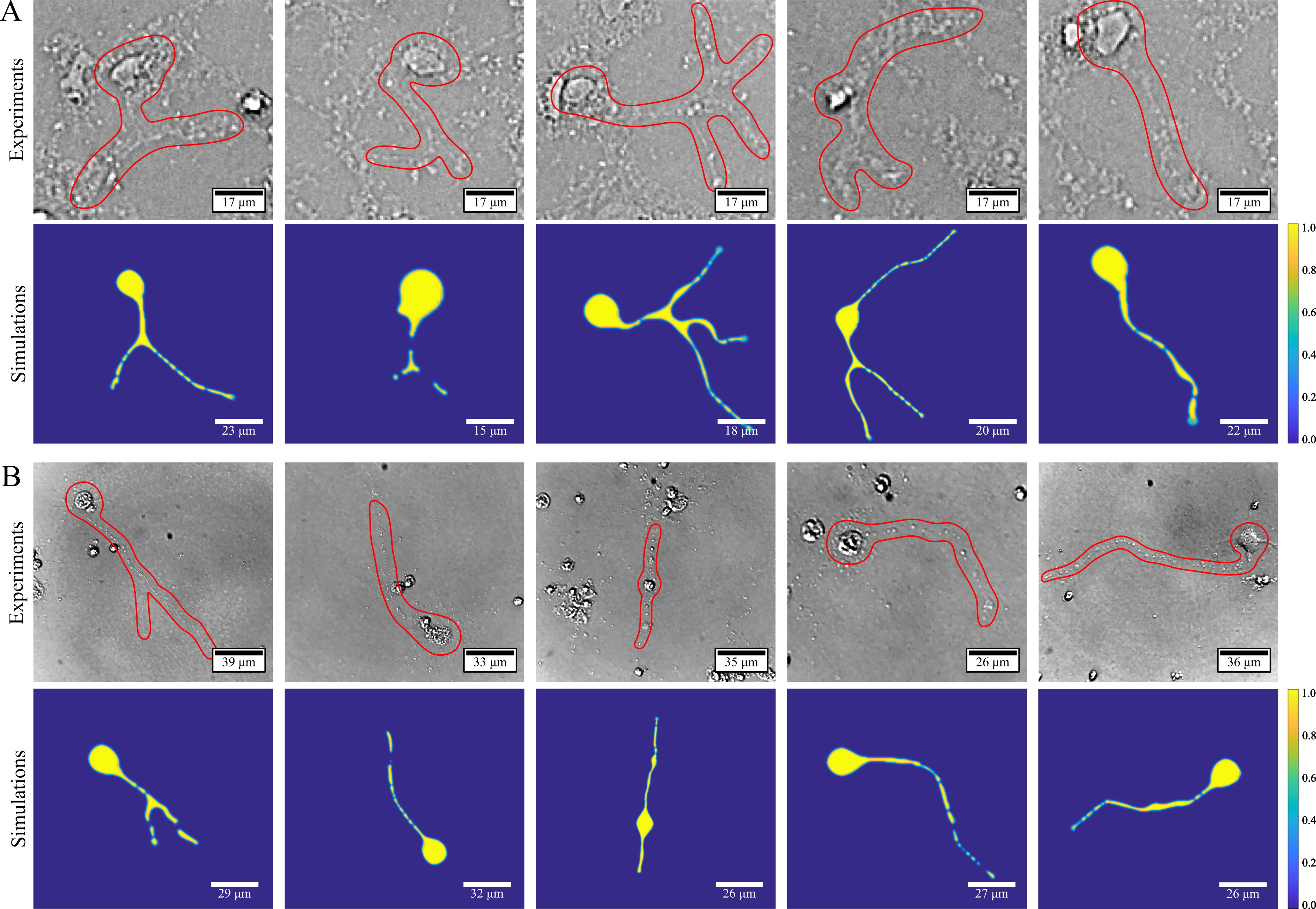
    }
    \caption{
        Additional comparison of experimental rat hippocampal neurons exhibiting dysfunction and simulated growth patterns. (A) Experimental (top) and model (bottom) neurons that experienced mechanical damage during culture.
        (B) Experimental (top) and model (bottom) neurons that exhibited neurite breaking and fragmentation from culture condition optimization during the first 3-4 days of culture \textit{in vitro}. These examples demonstrate the ability of the model to capture a wide array of growth patterns and subsequently qualitatively capture the neurite fragmentation and retraction observed in damaged and degenerating neurons.
        %(B) Configurations involving multiple neurons, capturing complex interactions and morphologies.
    }
    \label{fig: Exp comparison results - unhealthy}
\end{figure}

A crucial component of our NDDs model study involves comparing it with experimental observations. 
We apply the external-cue guided mechanism~\cite{qian2023biomimetic} to simulate the existence of external attractive cues in the extracellular matrix for several simulation cases and set the results against experimental neuron growth patterns (Fig.~\ref{fig: Exp comparison results - healthy}-\ref{fig: Exp comparison results - unhealthy}). 
By selectively placing external cues around the neuron, our model can effectively guide neurite outgrowth toward these external cues, allowing it to capture the complex and dynamic growth patterns observed in experimental cultures. 
These simulations explore how well our model reflects real-world neural growth behaviors.
%, thereby validating its effectiveness for studying NDDs.
% For healthy neuron comparison studies, we use human iPSC-derived neurons cultured at the Mayo Clinic to explore the characteristics of healthy neurons. 
This study compares simulation results with single neuron cultures to better understand the effect of model parameters on individual neurite morphology.
Although culturing denser neuron cultures is feasible, the densely overlapping neurite structures will significantly complicate neurite identifications and comparisons. 
Considering these limitations, efforts have been made to develop sparser neuron cultures to improve the clarity and reliability of neurite analysis.

\textbf{Comparison with healthy human iPSC-derived neurons.}
% Fig.~\ref{fig: Exp comparison results - healthy} showcases the effectiveness of our NDDs model in simulating neuron growth patterns observed in experimental settings.
First, we evaluate the model by comparing single-neuron growth patterns from day 1 to day 5 (Fig.~\ref{fig: Exp comparison results - healthy}A). 
Experimental images of human iPSC-derived neurons cultured over five days are used as the reference. 
We incorporate the external cue-guided mechanism by placing external cues around the soma to direct neurite outgrowth. 
These results indicate that the NDDs model can capture the observed neurite growth patterns when external cues are placed according to experimental culture images.
Next, we test the NDDs model with multiple neuron configurations (Fig.~\ref{fig: Exp comparison results - healthy}B). 
This setup involved comparing experimental images of multi-neuron growth with simulations where external cues were selectively placed to guide neurite outgrowth. 
The simulation outcomes closely aligned with the experimental images, demonstrating that the model can biomimetically capture complex growth behaviors in multi-neuron environments.
Although measuring the exact biological conditions and translating them into parameters in the phase field model present inherent challenges, this comparison showcases the effectiveness and potential of our NDDs model in capturing essential aspects of the neurodevelopmental process, highlighting its potential as a powerful tool for studying NDDs. 
These results evaluate the biomimetic capabilities and applicability of the NDDs model in simulating realistic neurodevelopmental processes, providing valuable insights for targeted therapeutic strategies in the future.

\textbf{Comparison with damaged and degenerating rat hippocampus neurons.}
% Utilizing the same external cue placement strategy~\cite{qian2023biomimetic}, we perform a detailed comparative analysis of two neuron growth cases with disorders by modifying $D_c$ and $c_{opti}$ and compared with rat hippocampal neuron cultures undergoing  neurite degeneration (Fig.~\ref{fig: Disorder singleCaseStudy}).
Utilizing the external cue placement strategy~\cite{qian2023biomimetic}, we perform a detailed comparative analysis of two neuron growth cases with disorders (Fig.~\ref{fig: Disorder singleCaseStudy}). 
This analysis involves modifying the diffusion coefficient $D_c$ and the optimal concentration $c_{opti}$. 
The results are compared with rat hippocampal neuron cultures exhibiting neurite fragmentation morphologies, characterized by observable severance of neurites followed by disintegration.
The NDDs model captures atrophy and retraction by adjusting $D_c$, which controls the diffusion of neurotrophin concentration, and $c_{opti}$, which sets the optimal neurotrophin concentration for neurodevelopmental processes. 
Based on experimental images of neuron growth undergoing damage and cell death (Fig.~\ref{fig: Disorder singleCaseStudy}A\&E).
Red outlines are empirically traced based on both experimental observations and accumulated expertise.
External cues are placed to guide healthy neurite patterns (Fig.~\ref{fig: Disorder singleCaseStudy}B\&F). 
Subsequently, we decrease the magnitude of $D_c$ from $6$ to $1$ to simulate reduced diffusion in neurotrophin concentration (Fig.~\ref{fig: Disorder singleCaseStudy}C\&G). 
This simulation reveals that neurites suffer significant atrophy and disconnections, showcasing a stark contrast to simulated healthy growth.
In the other comparison, we lowered $c_{opti}$ magnitude from $1$ to $0$ (Fig.~\ref{fig: Disorder singleCaseStudy}D\&H).
The results show that neurites not only experience significant atrophy but also demonstrate retraction behaviors, which are influenced by the cascading energy balance at the interface. 
This study highlights the critical influence of $D_c$ and $c_{opti}$ on pathological neuron growth affected by NDDs in our model.
These parameter adjustments show great potential to enable the NDDs model to simulate a variety of morphological abnormalities and provide insights into the dynamics of NDDs, aiding in the development of future targeted therapies.

In addition to the above two in-depth comparison studies, we apply our NDDs model to simulate a range of cases following the same approach (Fig.~\ref{fig: Exp comparison results - unhealthy}). 
The first and third rows in Fig.~\ref{fig: Exp comparison results - unhealthy} are the experimental images of neurons exhibiting degeneration and neurite fragmentation (in red outlines), following mechanical damage or culture condition optimization, respectively.
The second and fourth rows in Fig.~\ref{fig: Exp comparison results - unhealthy} are the corresponding NDDs model simulation results of NDDs neurite growth patterns.
Each simulation is able to simulate biomimetic growth behaviors under non-ideal conditions with atrophy and retraction. The ability of the NDDs model to qualitatively capture the neurite fragmentation and retraction observed during cell death and degeneration through biologically plausible parameter tuning highlights its potential for use in modeling NDDs. Future studies should calibrate parameters to individual disorders researchers wish to study.
% This detailed study allows us to fine-tune our computational model to capture actual experimental observations closely, ensuring that our NDDs model reflects real-world biological behaviors. 
% Furthermore, our approach highlights the great potential of computational models in advancing our understanding of complex NDDs pathogenesis and, therefore, assisting the development of effective treatments and therapeutic strategies for NDDs. 

\section{Conclusion and future work}
\label{sec: conclusion}

We present a novel computational model for NDDs, incorporating IGA, dynamic domain expansion and local refinement techniques into the phase field method.
The NDDs model demonstrates an uplift in computational performance compared to existing models and provides insights into complex neurite behaviors. 
We conclude that:
\begin{itemize}
    \item We have developed a PETSc-based NDDs model to utilize parallel processing with IGA and truncated T-splines. 
    This approach allows for accurate simulations of neuron growth processes and disorders, significantly enhancing computational efficiency without sacrificing accuracy.
    \item We introduced optimal neurotrophin concentration and degradation of neurotrophin to the phase field model and simulated NDDs neurite morphological transformations including retraction and atrophy, capturing the inverse relationship between neurotrophin levels and neuron survival. 
    %It uncovers the significant impact of neurotrophin on NDDs.
    \item The NDDs model incorporates dynamic domain expansion and targeted local refinements at the phase field interface. 
    This approach optimizes computational resources by expanding the domain dynamically based on the evolving neurite structures, ensuring high accuracy with reduced computational costs.
    \item We leveraged our computational NDDs model to investigate the intricate neurite morphological transformations associated with neural degeneration and NDDs, including retraction, atrophy, branching, and thickness variations. 
    \item The parameter study demonstrates the effect of each parameter on cell morphology resulting from the NDDs model with an external-cue guided mechanism. Qualitative comparisons with observed experimental neuron growth and eventual neurite breaking and fragmentation patterns, demonstrate its potential to advance the NDDs research.
    Our study unveils the functional roles of related factors and parameters within the model, offering critical hypotheses into underlying mechanisms of NDDs.
\end{itemize}

The model simulates neurite outgrowth morphologies through these novel techniques and expedites the study of NDDs. 
By capturing single- and multi-neuron dynamics, the model provides essential insights into the complex networks critical for understanding neurodevelopmental disorders. 
The parameter study reveals how the specific parameters influence neurite morphology and development within the model. For example,
how variations in parameters such as $k_{p75}$ and $D_c$ significantly alter neurite growth, affecting the thickness, branching, and survival of the neurites. 
% Our study identifies the parameters that most affect the neurodevelopmental process, laying a foundation for targeted research into NDDs. 
% Based on these findings, we aim to develop more accurate target therapeutic interventions and planning.
These findings, once validated in experimental models, may aid in developing more accurate target treatments and therapeutic planning.
In addition, our NDDs model can simulate biomimetic neurite outgrowth patterns utilizing an external-cue-guided mechanism when compared with experimental observations.
These additions will significantly broaden the potential of the NDDs model, positioning it as a promising computational tool for understanding the mechanisms behind disorders.
% This places the model as a crucial and effective computational tool for studying NDDs.

Because the current implementation is limited to 2D simulations, our future endeavors include extending the NDDs model to 3D implementation utilizing truncated hierarchical B-spline~\cite{pawar2016adaptive, li2023intracellular}. 
This improvement will facilitate a more precise representation of the complex 3D aspects of neuron morphology.
We plan to conduct more experiments with our collaborators to validate the model and ensure its accuracy and reliability in capturing neurodevelopmental processes.
To broaden our investigation, we intend to include unhealthy human iPSC-derived neurons currently being developed at the Mayo Clinic.
Despite parallelization and optimizations, the model still requires substantial computational resources, rendering it less practical for rapid and accurate predictions of neurite growth.
Furthermore, we will explore the integration of advanced machine learning techniques, including convolutional recurrent neural networks~\cite{shi_convolutional_2015}, and transformers~\cite{NIPS2017_3f5ee243}, to predict and analyze the time series evolution of neuron growth data. 
Incorporating machine learning models with physics-based simulations has proven successful for simple reaction-diffusion problems on 2D domain~\cite{li_reaction_2020} and complex neurite tree structures~\cite{li_deep_2021}.
In addition, physics-informed neural networks could significantly improve model performance~\cite{raissi_physics-informed_2019} for complex problems, including neuron traffic jams~\cite{li2023isogeometric}.
These advancements will significantly improve the model and broaden its potential in NDDs study.

\section*{Code and data availability}
The code and datasets generated and analyzed in this paper are accessible in the ``NNDs" GitHub repository. \url{https://github.com/CMU-CBML/NNDs} (DOI: \href{https://zenodo.org/records/12575160}{10.5281/zenodo.12575160}).
Correspondence and requests for code and data should be addressed to K.Q. or Y.J.Z.

\section*{Declaration of competing interest}
The authors declare no known competing financial interests or personal relationships that could have appeared to influence the work reported in this paper.

\section*{Acknowledgement}
K. Qian and Y. J. Zhang were supported in part by the NSF Grant CMMI-1953323. 
K. Qian, V. A. Webster-Wood, and Y. J. Zhang were supported in part by the Pennsylvania Infrastructure Technology Alliance (PITA) and Pennsylvania Manufacturing Innovation Program (PAMIP) grants. 
% K. Qian was also supported by Bradford and Diane Smith Graduate Fellowship. 
% In addition, V. A. Webster-Wood was supported in part by an NSF CAREER award ECCS-2044785. 
A. S. Liao and V. A. Webster-Wood were supported in part by an NSF CAREER award ECCS-2044785. 
A. S. Liao was also supported by NSF Graduate Research Fellowship Grant DGE-1745016 and Carnegie Mellon University Jean-Francois and Catherine Heitz Scholarship.
This work used RM-nodes on Bridges-2 Supercomputer at Pittsburgh Supercomputer Center~\cite{ecss,xsede} through allocation ID eng170006p from the Advanced Cyberinfrastructure Coordination Ecosystem: Services \& Support (ACCESS) program, which is supported by National Science Foundation grants \#2138259, \#2138286, \#2138307, \#2137603, and \#2138296.

\bibliographystyle{elsarticle-num}
\bibliography{reference}

\begin{thebibliography}{100}
\expandafter\ifx\csname url\endcsname\relax
  \def\url#1{\texttt{#1}}\fi
\expandafter\ifx\csname urlprefix\endcsname\relax\def\urlprefix{URL }\fi
\expandafter\ifx\csname href\endcsname\relax
  \def\href#1#2{#2} \def\path#1{#1}\fi

\bibitem{thapar2017neurodevelopmental}
A.~Thapar, M.~Cooper, M.~Rutter, Neurodevelopmental disorders, The Lancet Psychiatry 4~(4) (2017) 339--346.

\bibitem{tager1999neurodevelopmental}
H.~Tager-Flusberg, Neurodevelopmental disorders, MIT press, 1999.

\bibitem{fujitani2021pathophysiological}
M.~Fujitani, Y.~Otani, H.~Miyajima, Pathophysiological roles of abnormal axon initial segments in neurodevelopmental disorders, Cells 10~(8) (2021) 2110.

\bibitem{yamamoto2021genomic}
T.~Yamamoto, Genomic aberrations associated with the pathophysiological mechanisms of neurodevelopmental disorders, Cells 10~(9) (2021) 2317.

\bibitem{dugger_pathology_2017}
B.~N. Dugger, D.~W. Dickson, Pathology of neurodegenerative diseases, Cold Spring Harbor Perspectives in Biology 9 (2016) a028035.

\bibitem{brown_neurodegenerative_2005}
R.~C. Brown, A.~H. Lockwood, B.~R. Sonawane, Neurodegenerative diseases: an overview of environmental risk factors, Environmental Health Perspectives 113~(9) (2005) 1250--1256.

\bibitem{connor_role_1998}
B.~Connor, M.~Dragunow, The role of neuronal growth factors in neurodegenerative disorders of the human brain, Brain Research Reviews 27~(1) (1998) 1--39.

\bibitem{berg_new_1984}
D.~K. Berg, New neuronal growth factors, Annual Review of Neuroscience 7 (1984) 149--170.

\bibitem{elliott_motor_1996}
J.~L. Elliott, W.~D. Snider, Motor neuron growth factors, Neurology 47~(4 Suppl 2) (1996) 47S--53S.

\bibitem{datar2019roles}
A.~Datar, J.~Ameeramja, A.~Bhat, R.~Srivastava, A.~Mishra, R.~Bernal, J.~Prost, A.~Callan-Jones, P.~A. Pullarkat, The roles of microtubules and membrane tension in axonal beading, retraction, and atrophy, Biophysical Journal 117~(5) (2019) 880--891.

\bibitem{budday2015physical}
S.~Budday, P.~Steinmann, E.~Kuhl, Physical biology of human brain development, Frontiers in Cellular Neuroscience 9 (2015) 257.

\bibitem{wang2023multi}
S.~Wang, X.~Wang, M.~A. Holland, Multi-physics modeling and finite-element formulation of neuronal dendrite growth with electrical polarization, Brain Multiphysics 4 (2023) 100071.

\bibitem{hentschel_instabilities_1994}
H.~G.~E. Hentschel, A.~Fine, Instabilities in cellular dendritic morphogenesis, Physical Review Letters 73 (1994) 3592--3595.

\bibitem{krottje_mathematical_2007}
J.~K. Krottje, A.~Van~Ooyen, A mathematical framework for modeling axon guidance, Bulletin of Mathematical Biology 69 (2007) 3--31.

\bibitem{pearson_mathematical_2011}
Y.~E. Pearson, E.~Castronovo, T.~A. Lindsley, D.~A. Drew, Mathematical modeling of axonal formation {Part} {I}: {geometry}, Bulletin of Mathematical Biology 73 (2011) 2837--2864.

\bibitem{aeschlimann_biophysical_2001}
M.~Aeschlimann, L.~Tettoni, Biophysical model of axonal pathfinding, Neurocomputing 38-40 (2001) 87--92.

\bibitem{goodhill_predicting_2004}
G.~J. Goodhill, M.~Gu, J.~S. Urbach, Predicting axonal response to molecular gradients with a computational model of filopodial dynamics, Neural Computation 16~(11) (2004) 2221--2243.

\bibitem{maskery_growth_2004}
S.~M. Maskery, H.~M. Buettner, T.~Shinbrot, Growth cone pathfinding: a competition between deterministic and stochastic events, {BMC} Neuroscience 5~(22) (2004).

\bibitem{koene_netmorph_2009}
R.~A. Koene, B.~Tijms, P.~Van~Hees, F.~Postma, A.~De~Ridder, G.~J. Ramakers, J.~Van~Pelt, A.~Van~Ooyen, {NETMORPH}: a framework for the stochastic generation of large scale neuronal networks with realistic neuron morphologies, Neuroinformatics 7 (2009) 195--210.

\bibitem{cuntz_one_2010}
H.~Cuntz, F.~Forstner, A.~Borst, M.~Häusser, One rule to grow them all: a general theory of neuronal branching and its practical application, {PLoS} Computational Biology 6~(8) (2010) e1000877.

\bibitem{donohue_comparative_2008}
D.~E. Donohue, G.~A. Ascoli, A comparative computer simulation of dendritic morphology, {PLoS} Computational Biology 4~(6) (2008) e1000089.

\bibitem{torben-nielsen_context-aware_2014}
B.~Torben-Nielsen, E.~De~Schutter, Context-aware modeling of neuronal morphologies, Frontiers in Neuroanatomy 8 (2014) 92.

\bibitem{eberhard_neugen_2006}
J.~P. Eberhard, A.~Wanner, G.~Wittum, {NeuGen}: a tool for the generation of realistic morphology of cortical neurons and neural networks in 3{D}, Neurocomputing 70~(1-3) (2006) 327--342.

\bibitem{van_ooyen_independently_2014}
A.~van Ooyen, A.~Carnell, S.~de~Ridder, B.~Tarigan, H.~D. Mansvelder, F.~Bijma, M.~de~Gunst, J.~van Pelt, Independently outgrowing neurons and geometry-based synapse formation produce networks with realistic synaptic connectivity, {PLOS} {ONE} 9~(1) (2014) e85858.

\bibitem{otoole_physical_2008}
M.~O’Toole, P.~Lamoureux, K.~E. Miller, A physical model of axonal elongation: force, viscosity, and adhesions govern the mode of outgrowth, Biophysical Journal 94~(7) (2008) 2610--2620.

\bibitem{graham_mathematical_2006}
B.~P. Graham, A.~Van~Ooyen, Mathematical modelling and numerical simulation of the morphological development of neurons, {BMC} Neuroscience 7 (2006) S9.

\bibitem{li_deep_2021}
A.~Li, A.~Barati~Farimani, Y.~J. Zhang, Deep learning of material transport in complex neurite networks, Scientific Reports 11 (2021) 11280.

\bibitem{li2023isogeometric}
A.~Li, Y.~J. Zhang, Isogeometric analysis-based physics-informed graph neural network for studying traffic jam in neurons, Computer Methods in Applied Mechanics and Engineering 403 (2023) 115757.

\bibitem{takaki_phase_field_2015}
T.~Takaki, K.~Nakagawa, Y.~Morita, E.~Nakamachi, Phase-field modeling for axonal extension of nerve cells, Mechanical Engineering Journal 2~(3) (2015) 15--00063.

\bibitem{nella2022bridging}
K.~T. Nella, B.~M. Norton, H.-T. Chang, R.~A. Heuer, C.~B. Roque, A.~J. Matsuoka, Bridging the electrode--neuron gap: finite element modeling of \textit{in vitro} neurotrophin gradients to optimize neuroelectronic interfaces in the inner ear, Acta Biomaterialia 151 (2022) 360--378.

\bibitem{qian_modeling_2022}
K.~Qian, A.~Pawar, A.~Liao, C.~Anitescu, V.~Webster-Wood, A.~W. Feinberg, T.~Rabczuk, Y.~J. Zhang, Modeling neuron growth using isogeometric collocation based phase field method, Scientific Reports 12 (2022) 8120.

\bibitem{qian2023biomimetic}
K.~Qian, A.~S. Liao, S.~Gu, V.~A. Webster-Wood, Y.~J. Zhang, Biomimetic {IGA} neuron growth modeling with neurite morphometric features and {CNN}-based prediction, Computer Methods in Applied Mechanics and Engineering 417 (2023) 116213.

\bibitem{liao_quantitative_2022}
A.~Liao, W.~Cui, Y.~J. Zhang, V.~Webster-Wood, Quantitative evaluation of neuron developmental morphology \textit{in vitro} using the change-point test, Summer Biomechanics, Bioengineering and Biotransport Conference (2021).

\bibitem{van_ooyen_modeling_2003}
A.~van Ooyen, Modeling Neural Development, {MIT} Press, 2003.

\bibitem{zhang_challenges_2013}
Y.~J. Zhang, Challenges and advances in image-based geometric modeling and mesh generation, in: Image-Based Geometric Modeling and Mesh Generation, Springer, 2013.

\bibitem{zhang_geometric_2018}
Y.~J. Zhang, Geometric {Modeling} and {Mesh} {Generation} from {Scanned} {Images}, Chapman and Hall/{CRC}, 2016.

\bibitem{piegl1996nurbs}
L.~Piegl, W.~Tiller, The NURBS Book, Springer Science \& Business Media, 1996.

\bibitem{gordon_b-spline_1974}
W.~J. Gordon, R.~F. Riesenfeld, B-spline {Curves} and {Surfaces} (1974) 95--126.

\bibitem{hughes_isogeometric_2005}
T.~J. Hughes, J.~A. Cottrell, Y.~Bazilevs, Isogeometric analysis: {CAD}, finite elements, {NURBS}, exact geometry and mesh refinement, Computer Methods in Applied Mechanics and Engineering 194~(39-41) (2005) 4135--4195.

\bibitem{cottrell2009isogeometric}
J.~A. Cottrell, T.~J. Hughes, Y.~Bazilevs, Isogeometric analysis: toward integration of CAD and FEA, John Wiley \& Sons, 2009.

\bibitem{casquero_isogeometric_2016}
H.~Casquero, L.~Liu, Y.~J. Zhang, A.~Reali, H.~Gomez, Isogeometric collocation using analysis-suitable \uppercase{T}-splines of arbitrary degree, Computer Methods in Applied Mechanics and Engineering 301 (2016) 164--186.

\bibitem{sederberg2003t}
T.~W. Sederberg, J.~Zheng, A.~Bakenov, A.~Nasri, {T}-splines and {T-NURCCs}, ACM Transactions on Graphics 22~(3) (2003) 477--484.

\bibitem{sederberg2004t}
T.~W. Sederberg, D.~L. Cardon, G.~T. Finnigan, N.~S. North, J.~Zheng, T.~Lyche, {T}-spline simplification and local refinement, ACM Transactions on Graphics 23~(3) (2004) 276--283.

\bibitem{scott2012local}
M.~A. Scott, X.~Li, T.~W. Sederberg, T.~J. Hughes, Local refinement of analysis-suitable {T}-splines, Computer Methods in Applied Mechanics and Engineering 213 (2012) 206--222.

\bibitem{dokken2013polynomial}
T.~Dokken, T.~Lyche, K.~F. Pettersen, Polynomial splines over locally refined box-partitions, Computer Aided Geometric Design 30~(3) (2013) 331--356.

\bibitem{johannessen2014isogeometric}
K.~A. Johannessen, T.~Kvamsdal, T.~Dokken, Isogeometric analysis using {LR B}-splines, Computer Methods in Applied Mechanics and Engineering 269 (2014) 471--514.

\bibitem{kang2013modified}
H.~Kang, F.~Chen, J.~Deng, Modified {T}-splines, Computer Aided Geometric Design 30~(9) (2013) 827--843.

\bibitem{wei2022analysis}
X.~Wei, X.~Li, K.~Qian, T.~J. Hughes, Y.~J. Zhang, H.~Casquero, Analysis-suitable unstructured {T}-splines: multiple extraordinary points per face, Computer Methods in Applied Mechanics and Engineering 391 (2022) 114494.

\bibitem{liu2015weighted}
L.~Liu, Y.~J. Zhang, X.~Wei, Weighted {T}-splines with application in reparameterizing trimmed {NURBS} surfaces, Computer Methods in Applied Mechanics and Engineering 295 (2015) 108--126.

\bibitem{liu2015handling}
L.~Liu, Y.~J. Zhang, X.~Wei, Handling extraordinary nodes with weighted {T}-spline basis functions, Procedia Engineering 124 (2015) 161--173.

\bibitem{deng2008polynomial}
J.~Deng, F.~Chen, X.~Li, C.~Hu, W.~Tong, Z.~Yang, Y.~Feng, Polynomial splines over hierarchical {T}-meshes, Graphical Models 70~(4) (2008) 76--86.

\bibitem{evans2015hierarchical}
E.~Evans, M.~Scott, X.~Li, D.~Thomas, Hierarchical {T}-splines: analysis-suitability, {B}{\'e}zier extraction, and application as an adaptive basis for isogeometric analysis, Computer Methods in Applied Mechanics and Engineering 284 (2015) 1--20.

\bibitem{wei2015truncated}
X.~Wei, Y.~J. Zhang, T.~J. Hughes, M.~A. Scott, Truncated hierarchical {Catmull}--{Clark} subdivision with local refinement, Computer Methods in Applied Mechanics and Engineering 291 (2015) 1--20.

\bibitem{wei2016extended}
X.~Wei, Y.~J. Zhang, T.~J. Hughes, M.~A. Scott, Extended truncated hierarchical {Catmull--Clark} subdivision, Computer Methods in Applied Mechanics and Engineering 299 (2016) 316--336.

\bibitem{pawar2016adaptive}
A.~Pawar, Y.~J. Zhang, Y.~Jia, X.~Wei, T.~Rabczuk, C.~L. Chan, C.~Anitescu, Adaptive {FEM}-based nonrigid image registration using truncated hierarchical {B}-splines, Computers \& Mathematics with Applications 72~(8) (2016) 2028--2040.

\bibitem{li2020trivariate}
B.~Li, J.~Fu, Y.~J. Zhang, A.~Pawar, A trivariate {T}-spline based framework for modeling heterogeneous solids, Computer Aided Geometric Design 81 (2020) 101882.

\bibitem{li2019slicing}
B.~Li, J.~Fu, Y.~J. Zhang, W.~Lin, J.~Feng, C.~Shang, Slicing heterogeneous solid using octree-based subdivision and trivariate {T}-splines for additive manufacturing, Rapid Prototyping Journal 26~(1) (2019) 164--175.

\bibitem{casquero2017arbitrary}
H.~Casquero, L.~Liu, Y.~J. Zhang, A.~Reali, J.~Kiendl, H.~Gomez, Arbitrary-degree {T}-splines for isogeometric analysis of fully nonlinear {Kirchhoff--Love} shells, Computer-Aided Design 82 (2017) 140--153.

\bibitem{li2022modeling}
A.~Li, Y.~J. Zhang, Modeling intracellular transport and traffic jam in {3D} neurons using {PDE}-constrained optimization, Journal of Mechanics 38 (2022) 44--59.

\bibitem{li2022modeling_1}
A.~Li, Y.~J. Zhang, Modeling material transport regulation and traffic jam in neurons using {PDE}-constrained optimization, Scientific Reports 12 (2022) 3902.

\bibitem{petsc-user-ref}
S.~Balay, S.~Abhyankar, M.~F. Adams, S.~Benson, J.~Brown, P.~Brune, K.~Buschelman, E.~Constantinescu, L.~Dalcin, A.~Dener, V.~Eijkhout, J.~Faibussowitsch, W.~D. Gropp, V.~Hapla, T.~Isaac, P.~Jolivet, D.~Karpeev, D.~Kaushik, M.~G. Knepley, F.~Kong, S.~Kruger, D.~A. May, L.~C. McInnes, R.~T. Mills, L.~Mitchell, T.~Munson, J.~E. Roman, K.~Rupp, P.~Sanan, J.~Sarich, B.~F. Smith, S.~Zampini, H.~Zhang, H.~Zhang, J.~Zhang, {PETSc/TAO} users manual, Tech. Rep. ANL-21/39 - Revision 3.21, Argonne National Laboratory (2024).

\bibitem{petscsf2022}
J.~Zhang, J.~Brown, S.~Balay, J.~Faibussowitsch, M.~Knepley, O.~Marin, R.~T. Mills, T.~Munson, B.~F. Smith, S.~Zampini, The {PetscSF} scalable communication layer, IEEE Transactions on Parallel and Distributed Systems 33~(4) (2022) 842--853.

\bibitem{wei2017truncated}
X.~Wei, Y.~J. Zhang, L.~Liu, T.~J. Hughes, Truncated {T}-splines: {fundamentals} and {methods}, Computer Methods in Applied Mechanics and Engineering 316 (2017) 349--372.

\bibitem{karypis1998fast}
G.~Karypis, V.~Kumar, A fast and high quality multilevel scheme for partitioning irregular graphs, SIAM Journal on Scientific Computing 20~(1) (1998) 359--392.

\bibitem{gomez_accurate_2014}
H.~Gomez, A.~Reali, G.~Sangalli, Accurate, efficient, and (iso) geometrically flexible collocation methods for phase-field models, Journal of Computational Physics 262 (2014) 153--171.

\bibitem{schillinger_isogeometric_2015}
D.~Schillinger, M.~J. Borden, H.~K. Stolarski, Isogeometric collocation for phase-field fracture models, Computer Methods in Applied Mechanics and Engineering 284 (2015) 583--610.

\bibitem{takaki2014phase}
T.~Takaki, Phase-field modeling and simulations of dendrite growth, ISIJ International 54~(2) (2014) 437--444.

\bibitem{liao2023semi}
A.~S. Liao, W.~Cui, Y.~J. Zhang, V.~A. Webster-Wood, Semi-automated quantitative evaluation of neuron developmental morphology \textit{in vitro} using the change-point test, Neuroinformatics 21~(1) (2023) 163--176.

\bibitem{allen1979microscopic}
S.~M. Allen, J.~W. Cahn, A microscopic theory for antiphase boundary motion and its application to antiphase domain coarsening, Acta Metallurgica 27~(6) (1979) 1085--1095.

\bibitem{takaki2007phase}
T.~Takaki, M.~Asanishi, A.~Yamanaka, Y.~Tomita, Phase-field simulation during spherulite formation of polymer, Key Engineering Materials 345 (2007) 939--942.

\bibitem{eggleston2001phase}
J.~J. Eggleston, G.~B. McFadden, P.~W. Voorhees, A phase-field model for highly anisotropic interfacial energy, Physica D: Nonlinear Phenomena 150~(1-2) (2001) 91--103.

\bibitem{takaki2006two}
T.~Takaki, T.~Hasebe, Y.~Tomita, Two-dimensional phase-field simulation of self-assembled quantum dot formation, Journal of Crystal G rowth 287~(2) (2006) 495--499.

\bibitem{ren_controllable_2018}
B.~Ren, J.~Huang, M.~C. Lin, S.-M. Hu, Controllable dendritic crystal simulation using orientation field, Computer {Graphics} {Forum} 37~(2) (2018) 485--495.

\bibitem{mclean2004continuum}
D.~R. McLean, A.~van Ooyen, B.~P. Graham, Continuum model for tubulin-driven neurite elongation, Neurocomputing 58 (2004) 511--516.

\bibitem{mclean2004mathematical}
D.~R. McLean, B.~P. Graham, Mathematical formulation and analysis of a continuum model for tubulin-driven neurite elongation, Proceedings of the Royal Society of London. Series A: Mathematical, Physical and Engineering Sciences 460~(2048) (2004) 2437--2456.

\bibitem{graham2006dynamics}
B.~P. Graham, K.~Lauchlan, D.~R. Mclean, Dynamics of outgrowth in a continuum model of neurite elongation, Journal of Computational Neuroscience 20 (2006) 43--60.

\bibitem{van_ooyen_competition_2001}
A.~van Ooyen, B.~P. Graham, G.~J.~A. Ramakers, Competition for tubulin between growing neurites during development, Neurocomputing 38-40 (2001) 73--78.

\bibitem{kobayashi_modeling_1993}
R.~Kobayashi, Modeling and numerical simulations of dendritic crystal growth, Physica D: Nonlinear Phenomena 63~(3-4) (1993) 410--423.

\bibitem{song1997camp}
H.~Song, G.~Ming, M.~Poo, c{AMP}-induced switching in turning direction of nerve growth cones, Nature 388~(6639) (1997) 275--279.

\bibitem{bamji1998p75}
S.~X. Bamji, M.~Majdan, C.~D. Pozniak, D.~J. Belliveau, R.~Aloyz, J.~Kohn, C.~G. Causing, F.~D. Miller, The p75 neurotrophin receptor mediates neuronal apoptosis and is essential for naturally occurring sympathetic neuron death, The Journal of Cell Biology 140~(4) (1998) 911--923.

\bibitem{barrett2000p75}
G.~L. Barrett, The p75 neurotrophin receptor and neuronal apoptosis, Progress in Neurobiology 61~(2) (2000) 205--229.

\bibitem{meeker2014dynamic}
R.~Meeker, K.~Williams, Dynamic nature of the p75 neurotrophin receptor in response to injury and disease, Journal of Neuroimmune Pharmacology 9 (2014) 615--628.

\bibitem{meeker2015p75}
R.~B. Meeker, K.~S. Williams, The p75 neurotrophin receptor: at the crossroad of neural repair and death, Neural Regeneration Research 10~(5) (2015) 721--725.

\bibitem{marchetti2019fast}
L.~Marchetti, F.~Bonsignore, F.~Gobbo, R.~Amodeo, M.~Calvello, A.~Jacob, G.~Signore, C.~Schirripa~Spagnolo, D.~Porciani, M.~Mainardi, et~al., Fast-diffusing p75ntr monomers support apoptosis and growth cone collapse by neurotrophin ligands, Proceedings of the National Academy of Sciences 116~(43) (2019) 21563--21572.

\bibitem{krewson1996transport}
C.~E. Krewson, W.~M. Saltzman, Transport and elimination of recombinant human {NGF} during long-term delivery to the brain, Brain Research 727~(1-2) (1996) 169--181.

\bibitem{lu2005yin}
B.~Lu, P.~T. Pang, N.~H. Woo, The yin and yang of neurotrophin action, Nature Reviews Neuroscience 6~(8) (2005) 603--614.

\bibitem{piontek1999neurotrophins}
J.~Piontek, C.~C. Chen, M.~Kempf, R.~Brandt, Neurotrophins differentially regulate the survival and morphological complexity of human {CNS} model neurons, Journal of Neurochemistry 73~(1) (1999) 139--146.

\bibitem{huang2001neurotrophins}
E.~J. Huang, L.~F. Reichardt, Neurotrophins: roles in neuronal development and function, Annual Review of Neuroscience 24~(1) (2001) 677--736.

\bibitem{lyche1985knot}
T.~Lyche, E.~Cohen, K.~M{\o}rken, Knot line refinement algorithms for tensor product {B}-spline surfaces, Computer Aided Geometric Design 2~(1-3) (1985) 133--139.

\bibitem{buffa2010linear}
A.~Buffa, D.~Cho, G.~Sangalli, Linear independence of the {T}-spline blending functions associated with some particular {T}-meshes, Computer Methods in Applied Mechanics and Engineering 199~(23-24) (2010) 1437--1445.

\bibitem{giannelli2012thb}
C.~Giannelli, B.~J{\"u}ttler, H.~Speleers, Thb-splines: The truncated basis for hierarchical splines, Computer Aided Geometric Design 29~(7) (2012) 485--498.

\bibitem{giannelli2014strongly}
C.~Giannelli, B.~J{\"u}ttler, H.~Speleers, Strongly stable bases for adaptively refined multilevel spline spaces, Advances in Computational Mathematics 40 (2014) 459--490.

\bibitem{liunurbs}
L.~Liu, Y.~J. Zhang, X.~Wei, {NURBS} surface reparameterization using truncated {T}-splines, in: 23rd International Meshing Roundtable. London, UK. Oct. 12-15, 2014.

\bibitem{bentley1975multidimensional}
J.~L. Bentley, Multidimensional binary search trees used for associative searching, Communications of the ACM 18~(9) (1975) 509--517.

\bibitem{zhao2017apoe}
J.~Zhao, M.~D. Davis, Y.~A. Martens, M.~Shinohara, N.~R. Graff-Radford, S.~G. Younkin, Z.~K. Wszolek, T.~Kanekiyo, G.~Bu, {APOE} $\varepsilon$4/$\varepsilon$4 diminishes neurotrophic function of human i{PSC}-derived astrocytes, Human Molecular Genetics 26~(14) (2017) 2690--2700.

\bibitem{kawatani2023abca7}
K.~Kawatani, M.-L. Holm, S.~C. Starling, Y.~A. Martens, J.~Zhao, W.~Lu, Y.~Ren, Z.~Li, P.~Jiang, Y.~Jiang, et~al., Abca7 deficiency causes neuronal dysregulation by altering mitochondrial lipid metabolism, Molecular Psychiatry (2023) 1--11.

\bibitem{2018B-27System}
{Thermo Fisher Scientific}, {B-27 Plus Neuronal Culture System}, \url{https://assets.thermofisher.com/TFS-Assets/LSG/manuals/MAN0017319_B27_PlusNeuronalCultureSystem_UG.pdf} (2018).

\bibitem{gordon2021general}
J.~Gordon, S.~Amini, General overview of neuronal cell culture, Neuronal Cell Culture: Methods and Protocols (2021) 1--8.

\bibitem{yamamoto2016unidirectional}
H.~Yamamoto, R.~Matsumura, H.~Takaoki, S.~Katsurabayashi, A.~Hirano-Iwata, M.~Niwano, Unidirectional signal propagation in primary neurons micropatterned at a single-cell resolution, Applied Physics Letters 109~(4) (2016).

\bibitem{vogt2003micropatterned}
A.~K. Vogt, L.~Lauer, W.~Knoll, A.~Offenh{\"a}usser, Micropatterned substrates for the growth of functional neuronal networks of defined geometry, Biotechnology Progress 19~(5) (2003) 1562--1568.

\bibitem{haraguchi2012fabrication}
Y.~Haraguchi, T.~Shimizu, T.~Sasagawa, H.~Sekine, K.~Sakaguchi, T.~Kikuchi, W.~Sekine, S.~Sekiya, M.~Yamato, M.~Umezu, et~al., Fabrication of functional three-dimensional tissues by stacking cell sheets \textit{in vitro}, Nature Protocols 7~(5) (2012) 850--858.

\bibitem{gropp1996high}
W.~Gropp, E.~Lusk, N.~Doss, A.~Skjellum, A high-performance, portable implementation of the {MPI} message passing interface standard, Parallel Computing 22~(6) (1996) 789--828.

\bibitem{ecss}
N.~Wilkins-Diehr, S.~Sanielevici, J.~Alameda, J.~Cazes, L.~Crosby, M.~Pierce, R.~Roskies, An overview of the \uppercase{XSEDE} extended collaborative support program, in: High Performance Computer Applications - 6th International Conference, ISUM 2015, Vol. 595 of Communications in Computer and Information Science, 2016, pp. 3--13.

\bibitem{xsede}
J.~Towns, T.~Cockerill, M.~Dahan, I.~Foster, K.~Gaither, A.~Grimshaw, V.~Hazlewood, S.~Lathrop, D.~Lifka, G.~D. Peterson, R.~Roskies, J.~R. Scott, N.~Wilkins-Diehr, {XSEDE}: accelerating scientific discovery, Computing in Science \& Engineering 16~(5) (2014) 62--74.

\bibitem{diehl2016efficient}
S.~Diehl, E.~Henningsson, A.~Heyden, Efficient simulations of tubulin-driven axonal growth, Journal of Computational Neuroscience 41~(1) (2016) 45--63.

\bibitem{li2023intracellular}
A.~Li, Y.~J. Zhang, Intracellular material transport simulation in neurons using isogeometric analysis and deep learning, in: International Conference on Computational Science, Springer, 2023, pp. 486--493.

\bibitem{shi_convolutional_2015}
X.~Shi, Z.~Chen, H.~Wang, D.-Y. Yeung, W.-K. Wong, W.-C. Woo, Convolutional {LSTM} network: a machine learning approach for precipitation nowcasting, Advances in Neural Information Processing Systems 28 (2015).

\bibitem{NIPS2017_3f5ee243}
A.~Vaswani, N.~Shazeer, N.~Parmar, J.~Uszkoreit, L.~Jones, A.~N. Gomez, L.~U. Kaiser, I.~Polosukhin, Attention is all you need, Advances in Neural Information Processing Systems 30 (2017).

\bibitem{li_reaction_2020}
A.~Li, R.~Chen, A.~B. Farimani, Y.~J. Zhang, Reaction diffusion system prediction based on convolutional neural network, Scientific Reports 10 (2020) 3894.

\bibitem{raissi_physics-informed_2019}
M.~Raissi, P.~Perdikaris, G.~E. Karniadakis, Physics-informed neural networks: a deep learning framework for solving forward and inverse problems involving nonlinear partial differential equations, Journal of Computational Physics 378 (2019) 686--707.

\end{thebibliography}
\end{document}